\documentclass[journal,twocolumn,10pt]{IEEEtran}
\usepackage{graphicx}
\usepackage{verbatim}
\usepackage{upgreek}
\usepackage{amssymb,amsmath}
\usepackage{color}
\usepackage{epstopdf}
\usepackage{bm}
\usepackage{cite}
\usepackage{array,color}
\usepackage{algorithm}
\usepackage{algorithmicx}
\usepackage{algpseudocode}
\usepackage{amsmath}
\usepackage{amsfonts}
\usepackage{amsthm}
\usepackage{graphics}
\usepackage{epsfig}
\usepackage{soul}
\usepackage{booktabs}
\usepackage{multirow}
\soulregister\cite7
\soulregister\ref7
\usepackage{subfigure}
\usepackage{tabularx}
\usepackage{setspace}
\usepackage{makecell}
\usepackage{graphicx}
\usepackage{gensymb}
\renewcommand\appendix{\setcounter{secnumdepth}{3}}

\usepackage{etoolbox}
\newcommand{\tabincell}[2]{\begin{tabular}{@{}#1@{}}#2\end{tabular}} 
{}
\makeatletter
\patchcmd{\@makecaption}
{\\}
{.\ }
{}
{}
\makeatother

\ifCLASSINFOpdf
\else
\fi

\begin{document}

\title{\huge Latency Minimization Oriented Radio and Computation Resource Allocations for 6G V2X Networks with ISCC}

\author{Peng Liu,  Xinyi Wang,~\IEEEmembership{Member,~IEEE}, Zesong Fei,~\IEEEmembership{Senior~Member,~IEEE},\\Yuan Wu,~\IEEEmembership{Senior~Member,~IEEE}, Jie Xu,~\IEEEmembership{Fellow,~IEEE}, Arumugam Nallanathan,~\IEEEmembership{Fellow,~IEEE}
\vspace{-3mm}
	\thanks{Peng Liu,  Xinyi Wang, and Zesong Fei are with the School of Information and Electronics, Beijing Institute of Technology, Beijing 100081, China (E-mail: bit\_peng\_liu@163.com,  bit\_wangxy@163.com, feizesong@bit.edu.cn).}
		\thanks{Yuan Wu is with the State Key Laboratory of Internet of Things for Smart City, University of Macau, Macau SAR,  China, and also with Department of
		Computer and Information Science, University of Macau (E-mail: yuanwu@um.edu.mo).}
		\thanks{Jie Xu is with the School of Science and Engineering (SSE), the Shenzhen Future Network of Intelligence Institute (FNii-Shenzhen), and the Guangdong Provincial Key Laboratory of Future Networks of Intelligence, The Chinese University of Hong Kong (Shenzhen), Guangdong 518172, China (E-mail: xujie@cuhk.edu.cn). }
	   \thanks{Arumugam Nallanathan  is with the School of Electronic Engineering and Computer Science, Queen Mary University of London, E1 4NS London, U.K., and also with the Department of Electronic Engineering, Kyung Hee University, Yongin-si, Gyeonggi-do 17104, South Korea (E-mail: a.nallanathan@qmul.ac.uk). }
}

\maketitle
\begin{abstract}
Incorporating mobile edge computing (MEC) and integrated sensing and communication (ISAC) has emerged as a promising technology to enable integrated sensing, communication, and computing (ISCC) in the sixth generation (6G) networks. ISCC is particularly attractive  for vehicle-to-everything (V2X) applications, where vehicles  perform ISAC to sense the environment and simultaneously offload the sensing data to roadside base stations (BSs) for remote processing. In this paper, we investigate a particular ISCC-enabled V2X system consisting of multiple  multi-antenna BSs serving a set of  single-antenna vehicles, in which the vehicles perform their respective ISAC operations (for simultaneous sensing and offloading to the associated BS) over orthogonal sub-bands. With the focus on fairly minimizing the sensing completion latency for vehicles while ensuring the detection probability constraints, we jointly optimize  the allocations of radio resources (i.e., the sub-band allocation, transmit power control at vehicles, and receive beamforming at BSs) as well as computation resources at BS MEC servers. To solve the formulated complex mixed-integer nonlinear programming (MINLP) problem, we propose an alternating optimization algorithm. In this algorithm, we determine  the sub-band allocation via the branch-and-bound method, optimize the transmit power control  via successive convex approximation (SCA), and derive the receive beamforming  and computation resource allocation at BSs in closed form  based on generalized Rayleigh entropy and fairness criteria, respectively. Simulation results demonstrate that the proposed joint resource allocation design significantly reduces the maximum task completion latency among all vehicles. Furthermore, we also demonstrate  several  interesting trade-offs between the system performance and resource utilizations.

\vspace{2ex}
\end{abstract}

\section{Introduction}
\label{sec:intro}
The sixth generation  (6G) wireless networks are expected to support various emerging applications, and autonomous driving is among the most prominent ones. Towards this end, vehicles need to be equipped with sensing  capabilities to monitor the environment and make swift decisions for autonomous operation. To fulfill the high sensing, communication, and computation needs at vehicles, vehicle-to-everything (V2X) in 6G is particularly appealing, in which the vehicles and roadside base stations (BSs) can cooperate in sensing, communication, and computation \cite{xu2023}. Integrated sensing and communication (ISAC) and mobile edge computation (MEC) are becoming two important 6G technologies to support such operations in V2X networks. 

ISAC has been recognized by the International Telecommunication Union Radio communication Sector (ITU-R) as a foundational technology for 6G \cite{Kaushik2024}, enabling future wireless terminals and cellular BSs to support both high-speed data transmission and high-precision environmental sensing. Traditionally, radar sensing and communication are designed independently, using different hardware over non-overlapping  frequency bands. However, with the exponentially increasing number of wireless devices, the scarcity of spectrum resources has become a critical bottleneck for deploying separate  communication and sensing systems simultaneously. Fortunately, the similarities in hardware architecture, channel characteristics, and signal processing between sensing and communication present opportunities for the integration  of communication and radar sensing \cite{2022Liu}. In this context, ISAC has garnered increasing research attention. By jointly designing the sensing and communication functions on shared spectrum and hardware platforms, ISAC can significantly enhance both the spectral and energy efficiency with reduced hardware costs \cite{Wang2022tcom}. \textcolor{black}{Given these advantages,  one key potential application is to employ  ISAC technique in V2X networks to achieve high-precision sensing and provide high-throughput communication services for vehicles.}

On the other hand, the increasing volume of sensing data and the advancements of various signal processing and artificial intelligence (AI) algorithms  pose new challenges in meeting strict low-latency requirements in V2X scenarios. \textcolor{black}{For example, an autonomous vehicle must detect and respond to a sudden obstacle within tens of milliseconds to ensure driving safety \cite{ETSI}, which imposes stringent demands on both computation and communication capabilities.} To address this,  MEC has emerged as a promising solution to  process large amounts of data \cite{Xu2017} and support  advanced computation-intensive  sensing signal processing, such as spatial spectrum estimation \cite{radar5} and deep learning-based algorithms \cite{computing1}.  While vehicles may only have limited on-board computing power and energy, MEC enables them to offload such computation-intensive tasks to the network edge or cloud for remote processing, achieving high-accuracy and low-latency outcomes \cite{Zhangjsac2023}\cite{liu2024joint}. Consequently, integrating ISAC with MEC has been considered as a promising approach for integrating sensing, communication, and computing (ISCC) to support emerging V2X applications.

The emergence of ISCC raises new challenges in the joint radio and computation resource allocations, especially when the resources are distributed at different nodes and the sensing requests  and computation requests are space- and time-varying. For example, an aggressive power control scheme for sensing might cause severe interference towards communication and a bandwidth allocation scheme that benefits offloading rates may cause significant interference among sensing nodes. Therefore, proper resource allocation is important for realizing the full potential of ISCC.
\subsection{Related Works}
\subsubsection{Resource Allocation in ISAC Systems}
There have been extensive works studying resource allocation in ISAC systems from different perspectives on  time, bandwidth, and power resources. The authors of \cite{isacal1} proposed a dynamic frame structure based on for time-division  ISAC V2X systems, allowing all vehicles to coordinately allocate time ratios according to their different requests for sensing and communication, thus reducing communication delay.  For bandwidth and power allocations, the authors of \cite{isacal3} proposed an algorithm based on cyclic minimization, which jointly optimizes the subcarrier selection and power allocation to effectively reduce the power consumption while meeting the requirements on  sensing mutual information (MI) and communication rate.  \textcolor{black}{In \cite{isacal5}, dynamic power allocation in ISAC V2X networks was studied, with a stochastic programming problem formulated to maximize sensing MI. Simulation results showed improved trade-offs between sensing MI and communication rate.} Additionally, the authors of \cite{lyu2023} investigated trajectory optimization and power allocation in an unmanned aerial vehicle (UAV)-enabled ISAC system, formulating a novel sensing-constrained rate maximization problem to achieve  a better trade-off between sensing beampattern gain and communication rate. In \cite{wang2021tcom}, the authors investigated an ISAC network empowered by multiple UAVs, where power allocation, UAV positioning, and user association were jointly optimized to maximize network utility, subject to radar sensing constraints characterized by the Cramér-Rao lower bound (CRLB).
\subsubsection{Resource Allocation in MEC Systems}
Research on resource allocation in MEC systems have  primarily focused on coordinating bandwidth, power, and computation resources among terminals and MEC servers, with the objective of  optimizing the computation performance. The authors of \cite{Ding2020}  considered the  bandwidth resource allocation problem and proposed an iterative algorithm to minimize device energy consumption. Regarding power allocation, the authors of  \cite{yang2019} and \cite{MECal3} proposed  two iterative algorithms to reduce the minimum task delay of devices by optimizing the transmission power. Furthermore, the authors of \cite{Wu2024} and \cite{Ding2021} jointly optimized the computing offloading strategy, the allocation of computational resources at MEC servers, and the transmit power of devices to minimize the weighted sum latency. In addition, in \cite{kuang2024}, resource allocation for UAV-enabled MEC has been  investigated. An iterative algorithm has been  proposed for jointly optimizing UAV positioning, terminal transmission power, and MEC server computation resources, with the aim of reducing the energy consumption of UAVs.
\subsubsection{Resource Allocation in ISCC Systems}
There have only been a handful of prior works investigating the resource allocation for ISCC. The authors of \cite{liu2024joint} proposed a cloud–edge–device collaborative three-tier ISCC framework, where the overall processing latency of sensing tasks is reduced by optimizing the beamforming vectors and transmit power at the terminals. In \cite{ISCC4}, the authors studied sub-channel allocation in ISCC systems and proposed an iterative matching algorithm to enhance the overall network utility. \textcolor{black}{In \cite{ISCCiot1}, the authors explored a weighted optimization problem involving sensing MI and computation latency in ISCC systems, proposing a radio and computing resource allocation method based on the multi-agent actor-attention-critic model. Simulation results showed the trade-off between sensing MI and computation latency.} The authors of \cite{ISCC3} focused on ISCC-V2X networks, with the aim of improving the completion rate of cooperative sensing tasks via the joint optimization of bandwidth and edge computation resources. To examine the impacts of computation resources and power allocation on ISCC performance, the authors of \cite{ISCC2} developed a multi-objective optimization approach  to balance the energy consumption, offloading delay, and signal-to-interference-plus-noise ratio (SINR) for sensing. \textcolor{black}{In \cite{TITS1}, the authors studied a cooperative ISCC-V2X system and proposed a proximal policy optimization algorithm to optimize bandwidth, power, and computational resource allocation, with the aim of minimizing vehicle energy consumption. In \cite{ISCCiot2}, the authors addressed the power and computation resource allocation problem in ISCC-V2X and proposed a multi-agent deep deterministic policy gradient-based algorithm to reduce long-term average delay across vehicles.} The authors of \cite{NOMAMEC} investigated resource allocation in a non-orthogonal multiple access (NOMA)-aided ISCC system, considering both partial and binary offloading strategies. They jointly optimized the computational resources of MEC servers and the transmission power of terminals to enhance overall offloading rate. In \cite{Huang2024}, the authors investigated an MEC-aided ISAC system with short-packet transmissions, where ISAC beamforming matrices, short packet size, and computing resource allocation at the MEC server were jointly optimized to minimize the overall system energy consumption. Furthermore, the authors of \cite{UAVMEC} studied an ISCC-UAV system with multiple UAVs performing sensing tasks and data offloading, in which the energy consumption of UAVs is minimized via jointly optimizing UAV trajectories and power allocation.

\subsection{Motivation and Contributions}
Existing works  on ISCC resource allocation primarily  have focused on single-dimensional resource allocation design (e.g., bandwidth or power) \cite{liu2024joint} \cite{ISCC4}\cite{UAVMEC}, or joint allocation of power and computation resources \cite{ISCC3,ISCC2,NOMAMEC,ISCCiot1,ISCCiot2}\cite{bintcom}, with few studies exploring multi-dimensional resource allocation. Proper multi-dimensional resource allocation, including bandwidth, power, computation resource, and receive beamforming, can not only enhance the overall performance of the ISCC network, but also provide greater design flexibility for balancing multiple objectives (e.g., sensing accuracy, computing latency, and communication rate) and ensuring fairness in optimizing sensing and computing performance among terminals. 
	
Despite the above advantages, multi-dimensional resource allocation presents significant challenges. On one hand, the consideration of more resources leads to greater conflicts and coupling among communication, sensing, and computation. For example, both sensing and computation require higher power to achieve better echo SINR and lower task computation latency, respectively, and both functions also need effective sub-band allocation to mitigate interference. Balancing these conflicting resource demands is a key challenge in resource allocation design. On the other hand, multi-dimensional resource allocation increases the complexity of performance optimization. The joint optimization must account for inherent heterogeneity, such as varying computing capabilities of MEC servers and terminals, differing computation task demands, and diverse sensing requirements. Together with the complex interdependencies among resources, this adds another challenge.

Motivated by this, we investigate an ISCC V2X system consisting of a number of multi-antenna BSs and single-antenna vehicles, in which the vehicles perform their respective ISAC operations (for simultaneous sensing and offloading to the associated BS) over orthogonal sub-bands. We study the joint radio (e.g., the sub-band allocation, transmit power at vehicles, and receive beamforming at BSs) and computation resource allocation  to optimize the performance of sensing tasks, while ensuring vehicle detection performance for sensing. \textcolor{black}{Unlike existing ISCC resource allocation studies that integrate learning-based methods \cite{ISCCiot1, TITS1, ISCCiot2}, which require large amounts of data and long training periods, we propose a fully optimization-based resource allocation method, which avoids these challenges and ensures more predictable performance with lower computational complexity. The main contributions are summarized as follows.}
\begin{itemize}
	\item{Under the above setup, we formulate a joint optimization problem to minimize the maximum completion latency of sensing tasks across all vehicles under the constraints of detection probability and power budget,  through a design of joint radio and computation resource allocations. The sensing-performance-constrained latency minimization  is a mixed-integer nonlinear programming (MINLP) problem that is generally difficult to handle.} 
	\item {We propose an alternating optimization algorithm to address the MINLP problem. Firstly, we propose a novel branch-and-bound (BnB) algorithm  to decide the sub-band allocation. In particular, we first introduce a sensing-centric greedy  algorithm to generate a global bound for the sub-band allocation problem, and then use it along with sensing constraints to assist the branch pruning, thus effectively reducing the complexity. For the power control optimization, we employ the successive convex approximation (SCA) technique to reformulate the non-convex sub-problems as solvable convex forms. Subsequently, we derive the closed-form solutions for both  BSs' receive beamforming and MEC computation resource allocation based on the generalized Rayleigh entropy and fairness criteria, respectively. A comprehensive complexity analysis is conducted to show the efficiency of the proposed algorithm.}
	\item{Simulation results are presented to validate the performance of the proposed algorithm. It is shown that the proposed joint  resource allocation design significantly reduces the maximum sensing task completion latency for vehicles, as compared to benchmark schemes that only optimize the radio resource allocation or computation resource allocation. It is also shown  that the proposed joint design achieves  a better trade-off between target detection probability and task completion latency than other benchmark schemes with sensing-centric and computing-centric resource allocations. By evaluating the sensing task completion latency under different resource configurations, we also demonstrate the balance between the utilization of spectral and computation resources  while maintaining consistent latency performance.}
	
\end{itemize}

The remainder of this paper is organized as follows. Section II introduces the system model for the ISCC-enabled V2X network and formulates the associated optimization problem. In Section III, an efficient algorithm is developed to solve the joint resource allocation task. Section IV presents simulation results to validate the effectiveness of the proposed approach. Finally, Section V concludes the paper.

\textit{Notations:} The uppercase bold letters denote matrices, while lowercase bold letters represent vectors. The expectation operator is denoted by $\mathbb{E}[\cdot]$, and $\mathbb{C}$ denotes the set of complex numbers. For matrix operations, $\text{tr}(\cdot)$ indicates the trace, $[\cdot]^H$ the Hermitian (conjugate transpose), $[\cdot]^*$ the complex conjugate, and $[\cdot]^{-1}$ the inverse. The norm of a vector is written as $||\cdot||$. \textcolor{black}{A summary of key notations used in the paper is provided in Table I, where given constants are labeled as “Con”, derived constants or intermediate variables are labeled as “D-int”  and optimization variables are labeled as “Opt”.}

\begin{table}[!t]
	\renewcommand{\arraystretch}{1.05}
	\caption{Summary of Key Notations}
	\label{table_example}
	\centering
	\begin{tabular}{l l}
		\hline
		\bfseries Notation & \multicolumn{1}{c}{\bfseries Description}\\ 
		\hline
		$N$ & \tabincell{l}{Numbers of each BS's antennas (Con)}\\
		$K$ & \tabincell{l}{Numbers of vehicles (Con)}\\
		$M$ &  \tabincell{l}{Numbers of BSs (Con)}\\
		$K_m$ & \tabincell{l}{Number of vehicles served by BS $m$ (Con)}\\
		$L$ & \tabincell{l}{Number of sub-bands (Con)}\\
		$\mathbf{u}_k$ & \tabincell{l}{Receive beamforming vector for vehicle $k$ (Opt)} \\
		$\mathbf{A}$ & \tabincell{l}{Binary sub-band allocation martix (Opt)} \\
		$p_{k}$ & \tabincell{l}{Vehicle $k$'s transmission power  (dBm, Opt)}\\
		$B$ &  \tabincell{l}{The bandwidth of each sub-band (MHz, Con)} \\
		$\xi_k$ &  The RCS of vehicles $k$'s sensing target (Con) \\
		$P_{max}$ & \tabincell{l}{Vehicle's power budget (dBm, Con) }\\
		$R_{k,m}$ & \tabincell{l}{Achievable data rate from vehicle $k$ to  \\ BS $m$ (bits/s, D-int)} \\
		$b_k$ &   \tabincell{l}{Volume of radar data generated\\ by vehicle $k$ (bits/s, D-int)}  \\
		$e^L_k$ &   \tabincell{l}{CPU cycles required to process 1 bit for \\vehicle $k$ (cycles/bit, Con)}  \\
		$e^M_k$ &   \tabincell{l}{CPU cycles required to process 1 bit for \\MEC server (cycles/bit, Con)}  \\
		$f_{L,k}$ & \tabincell{l}{Vehicle $k$'s CPU frequency (cycles/s, Opt)}\\
		$f_{m,k}$ & \tabincell{l}{CPU frequency allocated to vehicle \\$k$ by MEC server $m$ (cycles/s, Opt)}\\
		$\kappa$ & \tabincell{l}{The coefficient determined by the hardware\\ structure (Con)}\\
		$\eta$ & \tabincell{l}{Ratio of the data volume offloaded to \\the overall data volume (Con)}\\
		$\gamma_k$ &   \tabincell{l}{Radar echo's SINR (bits/s, D-int)}  \\
		$S_{len}$ & \tabincell{l}{Maximum storage capacity of each layer for\\ BnB algorithm (Con)}\\
		\hline
	\end{tabular}
\end{table}

\section{System Model}
\label{sec:format}
\begin{figure}[!t]
	\centering
	\includegraphics[width=3.5in]{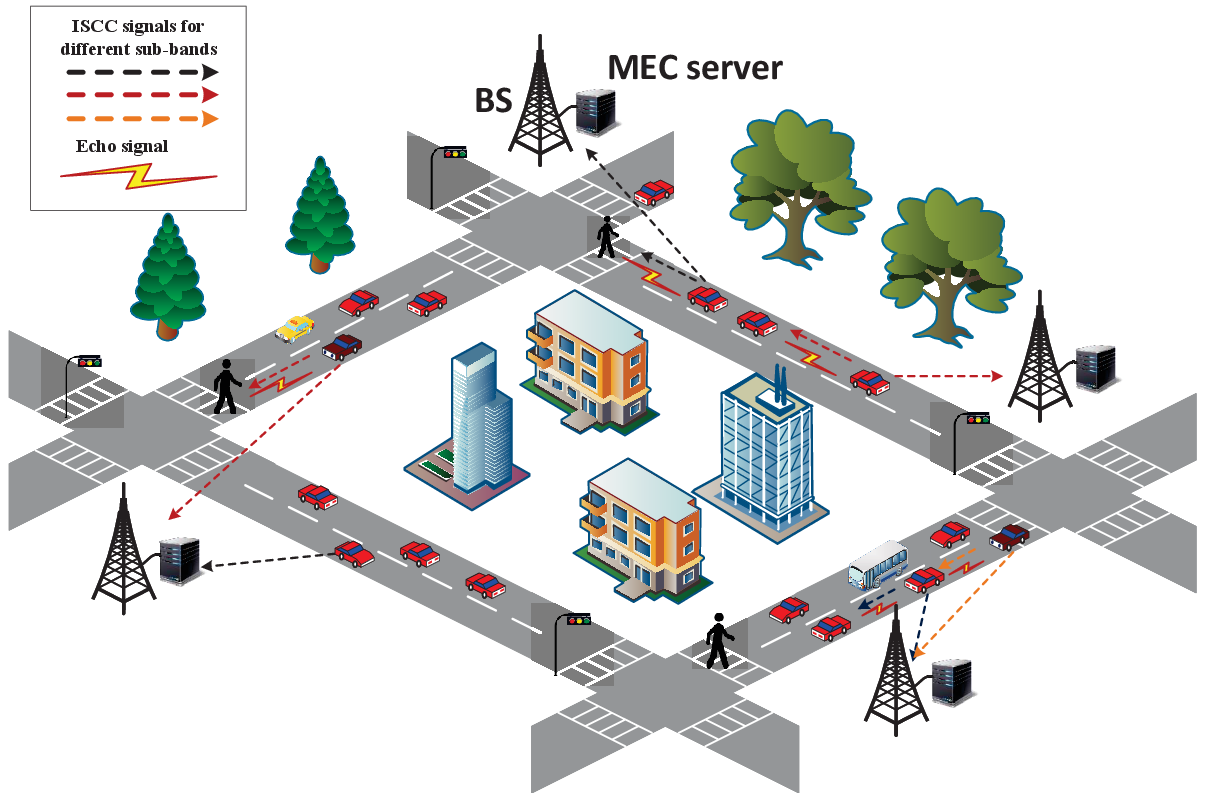}
	\caption{An illustration of ISCC V2X network.}
	\label{fig:sys}
\end{figure}

As shown in Fig.~\ref{fig:sys}, we consider an ISCC V2X network consisting of $M$ BSs, with each equipped with $N>1$ antennas and indexed as $\mathcal{M}=\{1,...,M\}$, and $K$ single-antenna vehicles\footnote{\textcolor{black}{The model can also be extended to the scenarios where each vehicle is deployed with  multiple antennas. In such cases, the transmit beamforming design must be carefully optimized to balance the performance of uplink offloading and sensing. Due to space limitations, we leave these considerations for future work.}} indexed as $\mathcal{K}=\{1,...,K\}$. Each BS is equipped with an MEC server to provide edge computing capabilities for vehicles. The vehicles transmit ISAC signals to simultaneously perform radar sensing and optionally offload the  computation-intensive sensing tasks  to the MEC servers\footnote{The BSs  can also act as data fusion centers, integrating the sensing data from multiple vehicles and returning the fused results to each vehicle for cooperative sensing. }. We assume that each  BS $m$ serves $K_m$ vehicles and denote the set of vehicles servered by BS $m$ as $\mathcal{K}_m$. \textcolor{black}{Here, we assume that $\mathcal{K}_m$'s are pre-determined by considering load balancing among different BSs}\footnote{\textcolor{black}{Numerous studies have addressed the user association problem in MEC systems, with a focus on balancing the communication and computation loads across BSs \cite{LB2022,LB2019}. Accordingly, this paper focuses on the resource allocation aspect, assuming that user association has been predetermined.}}. Due to the limited spectrum resources, we assume that all BSs reuse the same spectrum resources to serve their respectively associated vehicles. Specifically, the overall available bandwidth  is divided into $L$ orthogonal sub-bands, indexed as $\mathcal{L}=\{1,...,L\}$, where $L<K_m,\forall m$. To reduce the interference among different vehicles and vehicle-BS links, BSs cooperatively allocate spectrum  and power resources for its associated vehicles through a centralized control center with high computational capabilities \footnote{\textcolor{black}{We adopt an integrated approach of Frequency Division Multiple Access (FDMA) and Space Division Multiple Access (SDMA) due to their simplicity, and wide applications. The design principles obtained in this paper are extendable to other scenarios with  NOMA, Sparse Code Multiple Access (SCMA), or Reconfigurable Intelligent Surface (RIS)-assisted reuse techniques.}}. For clarity, we define a $L\times K$ binary sub-band allocation matrix as $\mathbf{A}$ with elements $\{a_{lk}\}$, where $l$ denotes the row index and $k$ denotes the  column index. If sub-band $l$ is allocated to vehicle $k$, it follows that $a_{lk}=1$; otherwise, $a_{lk}=0$.

\vspace{-2pt}  
\subsection{ISCC Signal Model}
  The channel vector from vehicle $k$ to BS $m$ over sub-band $l$ is denoted as $\mathbf{h}^l_{k,m }\in\mathbb{C}^{N \times 1}$. 
The  signal received at BS $m$ over sub-band $l$ can be expressed as
\begin{equation}
\label{yl} 
\color{black}\mathbf{r}^l_{k,m}=a_{lk}\mathbf{h}^{l}_{k,m}s_{k}+\sum_{i=1,i\neq k}^Ka_{li}\mathbf{h}^{l}_{i,m}s_{i} +\mathbf{n}^l_m,
\end{equation}
where $s_{k}$ denotes the signal transmitted from vehicle $k$ and $\mathbb{E}[s_ks_k^*]=p_k$, with $p_k$ being vehicle $k$'s transmission power. $\mathbf{n}^l_{m}\sim \mathcal{CN}(0, {\sigma_u}^2\mathbf{I}_N)$  denotes the additive white Gaussian noise (AWGN) at BS $m$ over sub-band $l$. The second term of (\ref{yl}) represents the co-channel interference. 

The achievable data rate (bit/s) from vehicle $k$ to BS $m$ is given by
\begin{equation}
\label{r1}
R_{k,m} = \sum_{l=1}^La_{lk}B\log_2\left(1+\frac{p_k\mathbf{u}^{H}_k\mathbf{h}^l_{k,m}(\mathbf{h}_{k,m}^{l})^H\mathbf{u}_k}{\mathbf{u}^{H}_k\mathbf{D}^l_k\mathbf{u}_k}\right),
\end{equation}
where $B$ denotes the bandwidth of each sub-band\footnote{\textcolor{black}{We consider a narrow-band system model to reflect practical V2X scenarios with limited spectrum and hardware constraints. Although wideband can improve both communication and sensing performance, it also requires more spectrum and adds hardware complexity.}}, $\mathbf{u}_k\in\mathbb{C}^{N \times 1}$ denotes the receive beamforming vector for vehicle $k$'s transmit signal,  and $\mathbf{D}^l_k=\sum_{i=1,i\neq k}^Ka_{li}p_i\mathbf{h}^l_{i,m}(\mathbf{h}^l_{i,m})^H +{\sigma^2_u\mathbf{I}_N}$.

In the sensing process, we assume that each vehicle performs target detection separately. To avoid collisions, the vehicles must detect the presence of targets within a minimum safe distance.  The received signals at vehicle $k$ over sub-band $l$, including the echo signal from the target of interest and interference from other vehicles allocated the same sub-band, can be expressed as\footnote{It should be noted that the vehicles operate in mono-static sensing, potentially leading to significant self-interference (SI). Nevertheless, it has been shown in \cite{2022Liu} that advanced SI cancellation techniques are able to achieve up to 100 dB SI suppression. Therefore, in this paper, we omit the effect of SI.} 
\begin{equation}
\color{black}{r}^l_{s,k}[w]=a_{lk} \alpha_k s_{k}[w]+\sum_{i=1, i \neq k}^K a_{li}{g}^{l}_{i,k}s_{i}[w]+{n}^l_r[w],
\end{equation}
where  $\alpha_k= \sqrt{\frac{{\rho_0}G_t}{{4\pi d_{t,k}}^{2}} \times \frac{{\xi}_kA_r}{{4\pi d_{t,k}}^{2}}}$ \cite{lptvt} with ${\xi}_k$ being the radar cross-section (RCS) of target to be sensed by vehicle $k$, $\rho_0$ being the path loss at the reference distance $d_0=1$m, $G_t$ and $A_r$ being the transmit antenna gain and the effective aperture of the receive antenna, and $d_{t,k}$ being the distance from vehicle $k$ to its sensing target. $g^l_{i,k}$ is the interfering channel from vehicle $i$ to vehicle $k$ over sub-band $l$. $w = 1,...,N_s$ denotes the time index, $N_s$ denotes the number of symbols accumulated for sensing, and  $n^l_r\sim \mathcal{CN}(0, {\sigma^2_r})$ is the AWGN at vehicles over sub-band $l$\footnote{\textcolor{black}{The sensing model is simplified to focus on key aspects relevant to sensing SINR, making the analysis more manageable. However, the proposed method remains applicable to more realistic sensing channel models (e.g., incorporating angle of arrival and clutter). In these cases, advanced techniques like transmit beamforming and cluster interference cancellation must be carefully designed to ensure optimal sensing performance. Introducing these techniques, however, will increase complexity and signal processing latency. Resource allocation for these more complex models will be explored in our future work.}}. 
After receiving echoes, vehicle $k$ executes target detection via hypothesis test, which can be formulated as
\begin{equation} \label{rsk}
\color{black}{r}^l_{s,k}[w]=\left\{\begin{array}{l}
 \displaystyle a_{lk}\alpha_k s_{k}[w]+\sum_{i=1, i \neq k}^K a_{li}{g}^l_{i,k}s_{i}[w]+{n}^l_r[w], \mathcal{H}_1, \\
 \displaystyle \sum_{i=1, i \neq k}^K a_{li}{g}^l_{i,k}s_{i}[w]+{n}^l_r[w], ~~\quad\quad\qquad~~~\mathcal{H}_0.
\end{array}\right.
\end{equation}
~~In (\ref{rsk}), $\mathcal{H}_1$ indicates that the target is present and $\mathcal{H}_0$ indicates that the target is absent. \textcolor{black}{By combining the signals from the $L $ sub-bands and using  the Neyman-Pearson criterion along with the generalized likelihood ratio test (GLRT),} the optimal detector is \cite{1998dec}
\begin{equation}
\color{black}T=\sum_{l=1}^L\sum_{w=1}^{N_s}|{r}^l_{s,k}[w]|^2 \stackrel{\mathcal{H}_0}{\underset{\mathcal{H}_1}{\lessgtr}} \psi,
\end{equation}
where $\psi$ is a threshold related to the constant false alarm probability and can be given by 
\begin{equation} \label{psi}
\psi= F_{\mathcal{X}_{2}^2}^{-1}\left(1-P_{F A}\right).
\end{equation} 
~~In (\ref{psi}), $P_{FA}$ is the acceptable false alarm probability and ${\mathcal {F}^{-1}_{\mathcal {X}_{2}^{2}\left ({\cdot}\right)}}$ is the inverse operation of the Chi-square cumulative distribution function (CDF). After matched filtering and whitening filtering, the asymptotic statistic of $T$  obeys the following distribution
\begin{equation}
\color{black}T \sim \begin{cases}  \mathcal{X}_{2}^2(\gamma_k): \frac{1}{2}  e^{-(T+\gamma_k) / 2}{\Gamma}_0(\sqrt{\gamma_kT}), & \mathcal{H}_1, \\\quad \mathcal{X}_{2}^2~~: \frac{1}{2}  e^{-T / 2}, & \mathcal{H}_0, \end{cases}
\end{equation}
where $\mathcal{X}_{2}^{2}(\gamma_k)$ represents the non-centrality \textcolor{black}{chi-square} distribution with degrees of freedom being two \textcolor{black}{and  ${\Gamma}_0(\cdot)$ denotes the Bessel function of the first kind}. $\gamma_k$ is the 
non-centrality parameter for $\mathcal{Xs}_{2}^2(\gamma_k)$,   and it is expressed as
\begin{equation} 
\gamma_k=\sum_{l=1}^L\frac{a_{lk}N_sp_k\alpha_k^2}{\sum_{i=1,i\neq k}^Ka_{li}p_i||{g}^l_{i,k}||^2 +{\sigma^2_r}}.
\end{equation}

Consequently, the asymptotic detection probability $P_D$ can be expressed as 
\begin{equation} \label{detec_prob}
{P_{D}} =P\left(T>\psi \mid \mathcal{H}_1\right)= 1 - {\mathcal {F}_{\mathcal {X}_{2}^{2}\left ({\gamma_k }\right)}}\left ({{\mathcal {F}_{\mathcal {X}_{2}^{2}\left ({\gamma_k }\right)}^{ - 1}\left ({{1 - {P_{FA}}} }\right)} }\right),
\end{equation}
where ${\mathcal {F}_{\mathcal {X}_{2}^{2}}}$ represents the \textcolor{black}{chi-square} CDF. It has been established in \cite{1998dec} that  $P_D$  monotonically increases with respect to  $\gamma_k$. We note that the expression for the non-centrality parameter $\gamma_k$ is the same as radar echo's SINR, indicating that higher SINR leads to better detection performance. Consequently, we constrain $\gamma_k$ at the minimum detection distance $d_{min}$ to be no smaller than a specific threshold\footnote{Notably, constraint (\ref{detec_prob2}) typically depends on the target's RCS and the interference channel. For the former, we set it to a fixed and small value to ensure a high detection probability even when the target's RCS is low (e.g., for pedestrians). For the latter, we can employ advanced channel estimation methods  used in vehicle-to-vehicle communication \cite{V2V} to obtain it.}, i.e., 
\begin{equation} \label{detec_prob2}
\gamma_{k\mid d_{t,k}=d_{min}}  \geq \Gamma_{d},  \forall k\in \mathcal{K},
\end{equation}
where $\Gamma_{d}$ is the sensing SINR threshold.

{\bf{Remark 1}:} \textit{\textcolor{black}{Due to the mobility of vehicles, the instantaneous Channel State Information (CSI) typically varies rapidly. Similar to \cite{channel2,channel1}, we assume that resource allocation is performed over a time block consisting of multiple slots, with a total duration on the order of several hundred microseconds. In this case, the large-scale fading, primarily determined by the vehicle’s position, is assumed to remain approximately constant within a time block. \textcolor{black}{Meanwhile, small-scale fading is assumed to be constant within each slot but varies slowly across different slots, with the correlation between slots depending on factors such as vehicle speed and Doppler shifts. Furthermore, we assume perfect CSI for each slot to analyze the best resource coordination gains for the ISCC system within a given time block. The robust design of scenarios with imperfect per-block CSI is left for future work.}}	} 

\subsection{Task Computation Model}

In the radar sensing process, the vehicles execute real-time echo signal processing and analysis. Based on the practical radar engineering implementation and radar detection model \cite{ISCC2}\cite{2006radar}, the volume of echoes for vehicle $k$ can be modeled as
\begin{equation}  \label{radar1}
b_k=\zeta f_sq_k,
\end{equation}
where $\zeta$  is a constant related to the configuration of radar systems, $f_s$ denotes the sampling frequency of vehicle $k$, and $q_k$ denotes the quantization bits for each sample. Furthermore, we define the sensing computing task of vehicle $k$ as \textcolor{black}{$D_k=(b_k,e^L_k)$}, where \textcolor{black}{$e^L_k$} is the task computation workload/intensity, which represents the per-bit computational load measured in central processing unit (CPU)  cycles~\cite{Xu2017}. 

We consider that the radar data
processing consists two stages. Initially, the vehicle performs local data preprocessing, which mainly involves the removal of unwanted signals \cite{radar} and data compression \cite{Ding2020}. During the radar sensing process, received signals often contain redundant information unrelated to target detection, such as ground clutter, radio frequency interference, and noise. Through signal preprocessing, we can improve data quality and enhance radar detection performance. The preprocessed data is then offloaded to the BS for further analysis \cite{radar3}. Finally, the BS transmits the processed results back to the vehicle. 

The latency and power consumption of local preprocessing at vehicle $k$ are respectively expressed as
\begin{equation}  \label{t1}
t_k^L=\frac{{b}_ke^L_k}{f_{L,k}}, ~ k\in \{1,\cdots,K\},
\end{equation}
and 
\begin{equation}  \label{e1}
p_k^L=\kappa(f_{L,k})^3, ~ k\in \{1,\cdots,K\},
\end{equation}
where $f_{L,k}$ is vehicle $k$'s CPU frequency  (cycles/s), $e^L_k$ is the computation intensity (cycles/bit) of preprocessing, and $\kappa$ is the coefficient determined by the hardware structure \cite{Ding2021}. 

The latency of sensing task offloading and MEC server processing for vehicle $k$  can be expressed as
\begin{equation}  \label{t1}
t_k^o=\frac{\eta{b}_k}{R_{k,m}}+\frac{\eta{b}_ke^M_k}{f_{m,k}}, ~\forall k\in \{1,\cdots,K\},
\end{equation}
where $\eta$ denotes the ratio of the amount of offloaded data to the original amount of data, $e^M_k$ is the computational intensity (cycles/bit) of MEC server processing, and $f_{m,k}$ is the CPU frequency allocated to vehicle $k$ by BS $m$. In practice, the computing capabilities of MEC server and vehicles are limited. Therefore, the CPU frequency needs to meet the following constraints:
\begin{align}
 \sum_{k=1}^{K_m}{f}_{m,k} \leq \hat{F}_m, \forall m,\\
 0\leq f_{L,k} \leq F_l,\forall k,
 \end{align}
 where $\hat{F}_m$ denotes the computational capacity of MEC server and $F_l$ denotes the maximum CPU frequency for vehicles. Consequently, vehicle $k$'s total task completion latency and power consumption can be expressed as  
\begin{align}
\label{p1}
&T_{k}=t_k^L+t_k^o=\frac{{b}_ke^L_k}{f_{L,k}}+\frac{\eta{b}_k}{R_{k,m}}+\frac{\eta{b}_ke^M_k}{f_{M,k}}, \\ 
&P_{k} = p_k+p_k^L=p_k+\kappa(f_{m,k})^3.
\end{align}

\begin{figure}[!t]
	\centering
	\includegraphics[width=2.9in]{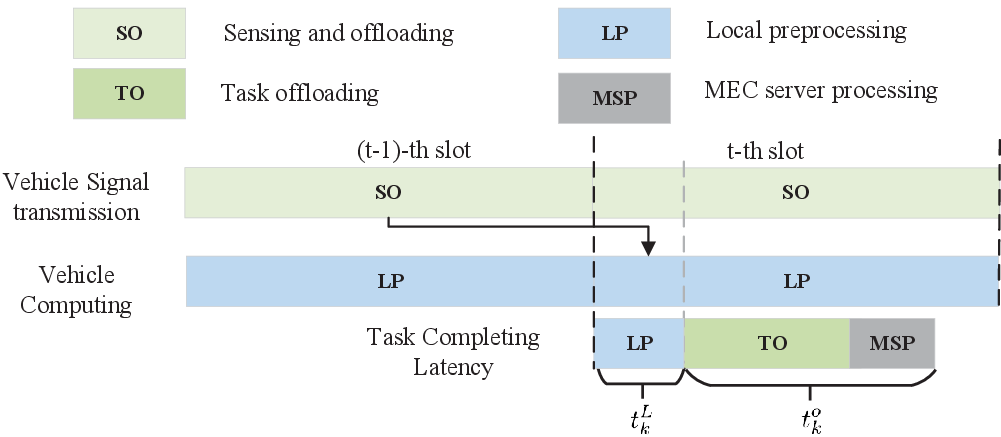}
	\caption{Slot structure of the ISCC V2X network.}
	\label{fig:frame}
\end{figure}

As the computational output is usually significantly smaller than the original sensing echoes, we do not consider the feedback delay from the BS to vehicles, \textcolor{black}{as in \cite{ISCC2,NOMAMEC,Ding2021}. Moreover, similar to these works, we only consider the power consumption of vehicles, while the computation power at the MEC server is not included, since MEC servers are typically connected to stable power sources with sufficient supply. To illustrate the composition of latency and power more clearly, we present the slot structure diagram in Fig.~\ref{fig:frame}. It is important to note that both computation and signal transmission significantly impact vehicles' power consumption \cite{NOMAMEC}, and therefore, we impose a joint constraint on them.}
\subsection{Problem Formulation}
From the above discussions, we observe that appropriate sub-band allocation and power control can effectively reduce intra-band interference, thereby improving both offloading rates and detection performance.  Furthermore, the offloading rate and the computation resource allocation directly influence the completion latency of sensing tasks. Therefore, in this paper, we focus on minimizing the completion latency of sensing tasks through  joint radio and computation resource allocation.  Specifically, we jointly optimize the sub-band allocation $\mathbf{A}$, power control $\mathbf{p}$, computation resources $\mathbf{f}_L,\mathbf{f}_M$, and BSs' receive beamforming $\mathbf{u}$ under the constraints of detection performance in (\ref{detec_prob2}), to minimize the maximum task completion latency for all vehicles $\max \limits_k ~ T_{k}$. This thus ensures the fairness among different vehicles. The joint optimization problem is formulated as
\begin{align} 
\label{prob29}
\min\limits _{\mathbf{A},\mathbf{p}, \mathbf{u},\mathbf{f}_L,\mathbf{f}_M} & ~\max \limits_k ~ T_{k}\\
s.t.~~~~
&\gamma_k \geq \Gamma_{d}, \forall k\in \mathcal{K}, \tag{\ref{prob29}a}\\
&{a_{lk}} \in \left\{ {0,1} \right\},~\forall l\in \mathcal{L},  k\in \mathcal{K},\tag{\ref{prob29}b}\\
&\sum_{l=1}^{L}{a}_{lk} \leq 1, \forall k\in \mathcal{K},\tag{\ref{prob29}c}\\
& \sum_{k=1}^{K_m}{f}_{m,k} \leq \hat{F}_m, \forall m,\tag{\ref{prob29}d}\\
&  0\leq f_{L,k} \leq F_l,\forall k \in \mathcal{K}, \tag{\ref{prob29}e}\\
& p_k+\kappa(f_{L,k})^3 \leq P_{max},\forall k\in \mathcal{K}, \tag{\ref{prob29}f}\\
& ||\mathbf{u}_k||=1,\forall k\in \mathcal{K}, \tag{\ref{prob29}g}
\end{align}
where  $P_{max}$ is the vehicle's maximum power budget, and $\{\mathbf{p},\mathbf{u},\mathbf{f}_L,\mathbf{f}_M\}\triangleq\{p_k,\mathbf{u}_k,f_{L,k},f_{m,k}\}^K_{k=1}$. Constraints (\ref{prob29}b) and (\ref{prob29}c) indicate that each vehicle is allocated with only one sub-band, while each sub-band can be assigned to multiple vehicles. From  problem ({\ref{prob29}}), we can see that the sub-band allocation matrix $\mathbf{A}$, transmit power $\mathbf{p}$, and BSs' receive beamforming $\mathbf{u}$ are coupled within the objective function and constraint (\ref{prob29}a). Moreover, problem ({\ref{prob29}}) is characterized as a MINLP problem due to the binary constraints in (\ref{prob29}b), making it particularly challenging to solve.

{\bf{Remark 2}:}\textit{ In problem ({\ref{prob29}}), there exist trade-offs between sensing and computing from multiple perspectives. Specifically, for the sensing process, it is preferable to allocate orthogonal channels to vehicles \textcolor{black}{that cause strong mutual interference, thereby increasing the SINR of the received sensing echoes. In contrast, for the computing and communication process, orthogonal channels should be assigned to vehicles closer to the BS to reduce vehicle-to-BS interference, thus improving the offloading rates.} \textcolor{black}{Additionally, to improve detection probability, vehicles tend to maximize their transmission power to achieve sufficient echo SINR. However, as shown in  constraint (\ref{prob29}f), increasing transmission power reduces the available power for local computing, which in turn increases task completion latency.}}

\section{Algorithm Design}
Due to the presence of coupled variables and the binary integer constraint, achieving the globally optimal solution of problem ({\ref{prob29}}) is challenging. 
Alternatively, we propose an alternating optimization algorithm to obtain a suboptimal but efficient solution. 
\textcolor{black}{Specifically, we decompose the original problem with five variables into four subproblems, namely, 
(1) sub-band allocation $\mathbf{A}$ using a low-complexity BnB approach; 
(2) joint transmit power $\mathbf{p}$ and local CPU frequency $\mathbf{f}_L$ optimization based on the SCA technique; 
(3) MEC servers' CPU frequency $\mathbf{f}_M$ optimization based on the  fairness criteria; and  (4)  BSs' receive beamforming $\mathbf{u}$ optimization based on the  generalized Rayleigh entropy.}

\subsection{Sub-band Allocation}
In this subsection, we optimize sub-band allocation matrix $\mathbf{A}$ with given $\{\mathbf{p},\mathbf{u},\mathbf{f}_L,\mathbf{f}_M\}$. By defining $w_k=\frac{1}{\eta{b}_k}$, problem ({\ref{prob29}}) can be rewritten as
\begin{align} 
\label{prob30}
\max\limits _{\mathbf{A}} & ~\min \limits_k ~ w_kR_{k,m}\\
s.t.
&~(\ref{prob29}\text{a}),(\ref{prob29}\text{b}),(\ref{prob29}\text{c}). \notag
\end{align}

\begin{figure}[!t]
	\centering
	\includegraphics[width=2.6in]{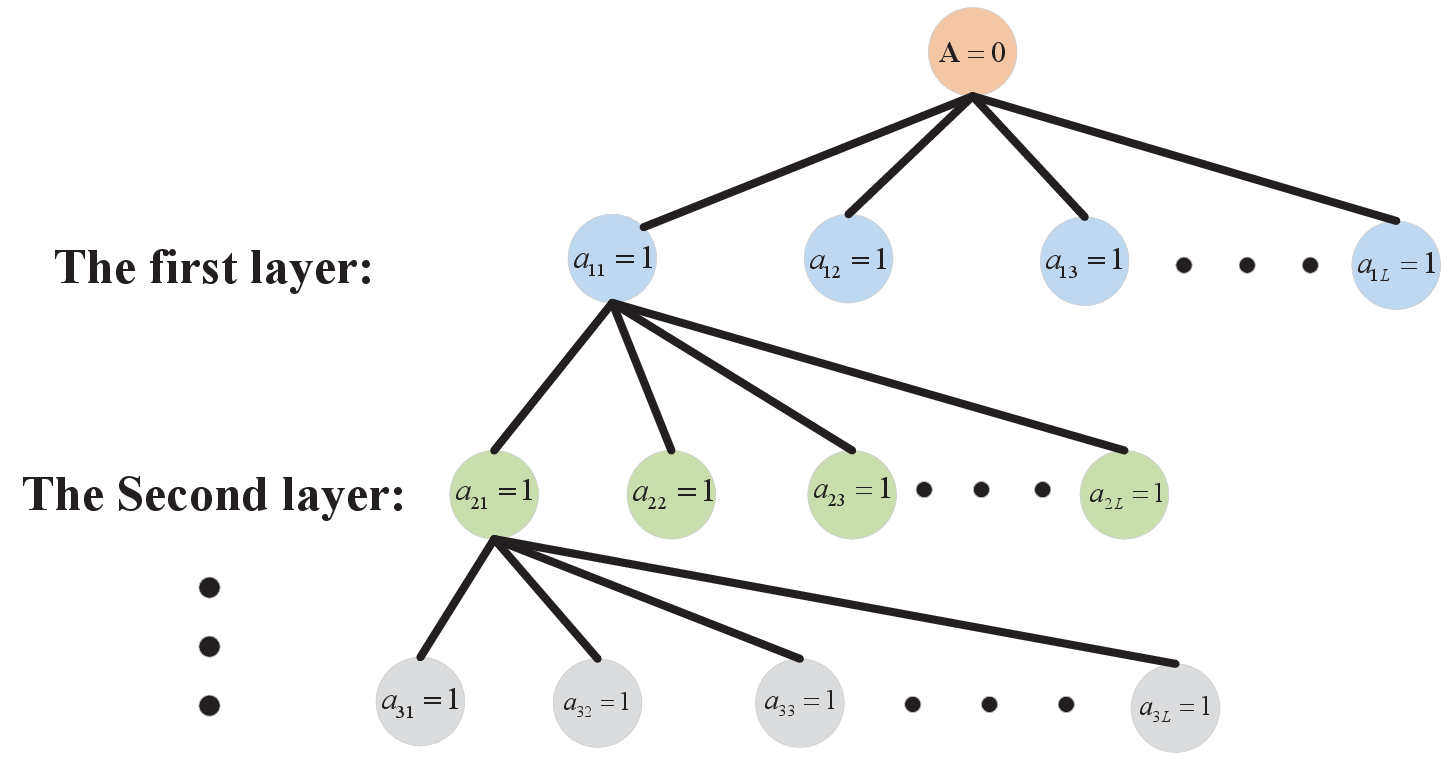}
	\caption{A search tree for the sub-band allocation matrix $\mathbf{A}$.}
	\label{fig:bnb}
\end{figure}

\subsubsection{Problem analysis}
Since there are a total of $K^L$ sub-band allocation strategies, it is almost impossible to solve subproblem ({\ref{prob30}}) through brute force search, especially when the number of vehicles and available sub-bands are large. A common approach is to relax the binary variables to continuous variables and then further employ convex optimization tools for solving. However, due to the form of mutual interference in the objective function and constraint ({\ref{prob29}}a), problem ({\ref{prob30}}) remains non-convex even after relaxation. To this end, we propose  a BnB  algorithm to directly solve the problem.

Firstly, as illustrated in Fig.~\ref{fig:bnb}, we model the search space for the sub-band allocation matrix $\mathbf{A}$ as an $L$-ary tree. Each leaf node within this $L$-ary tree represents a specific sub-band allocation scheme. We utilize a layer-by-layer method to allocate sub-bands to different vehicles. Initially, in the first layer, the $l$-th branch of the root node corresponds to exclusively assigning the $l$-th sub-band to vehicle $\mathit{1}$, while the other vehicles remain unassigned. Then, in the $k$-th layer,  sub-bands are allocated to the $k$-th vehicle, building upon the allocation completed for the first $k-1$ vehicles. 

This tree search problem can be optimally addressed using the classic BnB algorithm \cite{bnb1}. However, the traditional BnB method is highly complex. In the worst case, its complexity is comparable to that of a brute-force approach. Consequently, we introduce a sensing-centric greedy allocation algorithm (i.e., Algorithm 1) to  generate a global upper bound that facilitates the pruning process. Building on this, we further propose a streamlined and low-complexity BnB algorithm  (i.e., Algorithm 2).

\subsubsection{Greedy allocation algorithm}

The sensing-centric optimization problem with the aim of maximizing the minimum SINR of vehicles' sensing echoes is formulated as
\begin{align} 
\label{prob31}
\max\limits _{\mathbf{A}} & ~\min \limits_k ~ \gamma_k\\
s.t.
&~(\ref{prob29}\text{b}),(\ref{prob29}\text{c}). \notag
\end{align}

 To assess the level of interference between different vehicles, we further define the interference metric between vehicle $i$ and vehicle $j$ as $I_{i,j} =p_i||g_{i,j}||^2$. 

We begin by selecting two vehicles with the highest interference metric, \textcolor{black}{which represent the pair with the strongest mutual interference, since allocating orthogonal sub-bands to this pair first would maximize the benefit of interference mitigation.} Next, we sequentially choose the next $L-2$ vehicles that have the highest interference metrics with the already selected vehicles, assigning each vehicle to one of the remaining orthogonal sub-bands. To determine the sub-band allocation order for the remaining vehicles, we identify the next vehicle index to be assigned based on the following criteria:
\begin{equation} \label{select1}
\widetilde{k}=\arg \max _k \frac{1}{(L-1)}\left(\sum_{l=1}^L \sum_{j \in \mathcal{\widetilde{K}}_l} I_{k, j}-\min _l \sum_{j \in \mathcal{\widetilde{K}}_l} I_{k, j}\right),
\end{equation}
where $\mathcal{\widetilde{K}}_l$ denotes the set of vehicles that  have been allocated with the $l$-th sub-band. $\sum_{l=1}^L \sum_{j \in \mathcal{\widetilde{K}}_l} I_{k, j}$  denotes the total interference from vehicle $k$ to other vehicles when vehicle $k$ is assigned with different sub-bands. $\min _l \sum_{j \in \mathcal{\widetilde{K}}_l} I_{k, j}$ denotes the minimum interference metric introduced when vehicle $k$ is assigned different channels. Here,  (\ref{select1}) implies that we tend to select the vehicle that would introduce the maximum average interference if it is not assigned to the sub-band with the minimum interference metric. The sub-band allocated to vehicle $\widetilde{k}$ is
\begin{equation} \label{select2}
\widetilde{l}=\min _l \sum_{j \in \mathcal{\widetilde{K}}_l} I_{k, j}.
\end{equation}

In summary,  we summarize the detailed process of the greedy allocation algorithm in Algorithm 1. Under this scheme, the minimum offloading rate among vehicles  is set as the global lower bound $B_u$. Next,  we introduce the low-complexity BnB algorithm for solving problem (\ref{prob30}). 

\renewcommand{\algorithmicrequire}{\textbf{Input:}}
\renewcommand{\algorithmicensure}{\textbf{Output:}}
\begin{algorithm}[t] 
	\setstretch{1.0}
	\caption{Greedy allocation algorithm for Solving (\ref{prob31}) }
	\begin{algorithmic}[1] 
		\Require   Transmit power $\mathbf{p}$, local CPU frequency $\mathbf{f}_L$, MEC server CPU frequency $\mathbf{f}_M$,  sensing threshold $\Gamma_{d}$, BSs' receive beamforming $\mathbf{u}$.
		
		\State {Select the two vehicles corresponding to the highest  interference metric.}
		\Repeat
		\State {Select the vehicle that has the highest interference metric with the already selected vehicles.}
		\Until  {$L$ vehicles are selected.}
		\State {Allocate $L$ sub-bands to the $L$ vehicles.}
		\Repeat
		\State {Select the next vehicle  $\widetilde{k}$ to be allocated based on (\ref{select1}).}
		\State {Select the sub-band $\widetilde{l}$ allocated to the vehicle based on (\ref{select2}).}   
		\State {Add vehicle $k$ to the set $\mathcal{\widetilde{K}}_l$.}
		\Until {All vehicles have been assigned sub-bands.}
		\Ensure Sub-band allocation matrix $\mathbf{A}$.
	\end{algorithmic}
\end{algorithm}
\subsubsection{Low-complexity BnB algorithm}
\textcolor{black}{For the search tree depicted in Fig.~\ref{fig:bnb}, we adopt a breadth-first search approach to solve problem (\ref{prob30}), performing branching and bounding operations at each layer based on the global lower bound derived from problem (\ref{prob31}).} Specifically, at the first layer, we calculate the minimum offloading rate for each node as
\begin{equation} \label{bound}
B_d = \min\limits_{l} \min\limits_{{k\in\mathcal{\widetilde{K}}_l}}  ~ w_kB\log_2\left(1+\frac{p_k\mathbf{u}^{H}_k\mathbf{h}^l_{k,m}(\mathbf{h}^l_{k,m})^H\mathbf{u}_k}{\mathbf{u}_k\widehat{\mathbf{D}}^l_k\mathbf{u}^H_k}\right),
\end{equation}
where $\widehat{\mathbf{D}}^l_k=\sum_{i=1,i\neq k}^K\widehat{a}_{li}p_i\mathbf{h}^l_{i,m}(\mathbf{h}^l_{i,m})^H +{\sigma^2_u\mathbf{I}_N}$ with $\{\widehat{a}_{li}\}_{\forall l,i}$ being the sub-band allocation of the current node. We compare $B_d$ of each node with the global lower bound $B_u$ obtained by the greedy allocation algorithm. If $B_d<B_u$ or constraints in (\ref{prob29}a) are not satisfied, i.e., $\sum_{l=1}^L\frac{\widehat{a}_{lk}N_sp_k\alpha_k^2}{\sum_{i=1,i\neq k}^K\widehat{a}_{li}p_i||{g}^l_{i,k}||^2 +{\sigma^2_r}} < \Gamma_{d}, \forall k\in \mathcal{K}$, a pruning operation is performed. Subsequently, the remaining nodes are sorted in a descending order based on their minimum offloading rate, and the branching operation is repeated. To further reduce the complexity of the algorithm, we limit the maximum storage width of each layer to $S_{len}$, meaning that only the largest $S_{len}$ nodes are preserved for branching to the next layer. This process of recalculating boundaries and pruning is repeated until the final layer,  where each vehicle is assigned a sub-band. The node with the highest minimum offloading rate is then selected as the final sub-band allocation scheme. The detailed BnB algorithm is summarized in Algorithm 2\footnote{\textcolor{black}{In practical network deployments, the implementation of Algorithm 1 and Algorithm 2 requires the exchange of information between the centralized control center, BSs, and vehicles. First, the BS collects the vehicle's CSI based on pilot signals and uploads it to the control center. Then, the control center optimizes the sub-band allocation and sends the results back to both the BS and the corresponding vehicles. Finally, the vehicles perform sensing and data offloading on the assigned sub-band.}}.

\subsection{Power Control and Local Computation Resource Allocation}
We then optimize the transmit power  $\mathbf{p}$ and  vehicles' local CPU frequency $\mathbf{f}_L$ with  given $\{\mathbf{A},\mathbf{u},\mathbf{f}_M\}$. In this case, problem (\ref{prob29}) is reformulated as 
\begin{align} 
\label{prob32}
\min\limits _{\mathbf{p}, \mathbf{f}_L} & ~\max \limits_k ~ \frac{{b}_ke^L_k}{f_{L,k}}+\frac{\eta{b}_k}{R_{k,m}}\\
s.t.
&~~~(\ref{prob29}\text{e}),(\ref{prob29}\text{f}). \notag
\end{align}

By introducing slack variables $\mu_1$ and $\mu_2$, problem (\ref{prob32}) can be equivalently rewritten as
\begin{align} 
\label{prob33}
\min\limits _{\mathbf{p}, \mathbf{f}_L} & ~\mu_1+\mu_2\\
s.t.~
& \frac{{b}_ke^L_k}{f_{L,k}} \leq \mu_1, \forall k\in \mathcal{K},\tag{\ref{prob33}a}\\
& \frac{\eta{b}_k}{R_{k,m}} \leq \mu_2, \forall k\in \mathcal{K},\tag{\ref{prob33}b}\\
&(\ref{prob29}\text{e}),(\ref{prob29}\text{f}). \notag
\end{align}

\renewcommand{\algorithmicrequire}{\textbf{Input:}}
\renewcommand{\algorithmicensure}{\textbf{Output:}}
\begin{algorithm}[t] 
	\setstretch{1}
	\caption{Low-complexity BnB algorithm for Solving (\ref{prob30}) }
	\begin{algorithmic}[1] 
		\Require    Transmit power $\mathbf{p}$, local CPU frequency $\mathbf{f}_L$, MEC server CPU frequency $\mathbf{f}_M$, sensing threshold $\Gamma_{d}$, BSs' receive beamforming $\mathbf{u}$.
		\State {Calculate the global lower bound $B_u$ based on the solution from Algorithm 1.}
		\State {Initialize $i=1$.}
		\Repeat
		\State {Based on (\ref{bound}), calculate the minimum offloading rate $B_d$ for each node's sub-band allocation scheme at the $i$-th layer.}
		\State {Remove nodes that do not satisfy constraints (\ref{prob29}a) or $B_d<B_u$ and sort the minimum offloading rate $B_d$ of the remaining nodes at the $i$-th layer in descending order. }   
		\State {Retain $S_{len}$ nodes with the higher minimum offloading rate and branch them to the next layer.}
		\State {Update  $i = i+1$.}
		\Until {$i=K+1$. }
		\State {Select the sub-band allocation scheme corresponding to the node with the highest lower bound.}
		\Ensure the sub-band allocation matrix $\mathbf{A}$.
	\end{algorithmic}
\end{algorithm}

It is observed that problem (\ref{prob33}) remains challenging to solve due to the non-convex constraint (\ref{prob33}b). To make it tractable, we  introduce auxiliary variables $r_k, c_{1,k}$, and  $c_{2,k}, \forall k\in \mathcal{K}$. With these variables, we can  reformulate constraint (\ref{prob33}b) as
\begin{equation}
\label{eq1}
\frac{\eta{b}_k}{Br_{k}} \leq \mu_2, \forall k\in \mathcal{K},
\end{equation}
\begin{equation}
\label{eq2}
\frac{c_{1,k}}{c_{2,k}}\geq 2^{r_k}-1,  \forall k\in \mathcal{K}, 
\end{equation}
\begin{equation}
\label{eq3}
\sum_{l=1}^La_{lk}p_k\mathbf{u}_k^{H}\mathbf{h}^l_{k,m}(\mathbf{h}^l_{k,m})^H\mathbf{u}_k\geq c_{1,k},  \forall k\in \mathcal{K}, 
\end{equation}
 \begin{equation}\small
 \label{eq4}
 \sum_{l=1}^L\sum_{i=1,i\neq k}^Ka_{li}p_i\mathbf{u}^{H}_k\mathbf{h}^l_{i,m}(\mathbf{h}^l_{i,m})^H\mathbf{u}_k +\mathbf{u}^H_k{\sigma^2_u}\mathbf{I}_N\mathbf{u}_k\leq c_{2,k},  \forall k\in \mathcal{K}.
 \end{equation}
 
 Due to the fractional term on the left side of the inequality, constraint (\ref{eq2}) remains non-convex. Next, we develop an iterative algorithm using the SCA technique to transform this constraint into a convex form. By introducing auxiliary variables $v_{1,k}$, $v_{2,k}$, and $v_{3,k}$, constraint (\ref{eq2}) is decomposed as
 \begin{align}
&c_{1,k}\geq e^{\upsilon_{1,k}},~ \upsilon_{1,k}-\upsilon_{2,k}\geq \upsilon_{3,k},~~\forall k\in \mathcal{K},  \tag{\ref{eq2}a} \\
&c_{2,k}\leq e^{\upsilon_{2,k}},~ 2^{r_k}-1\leq e^ {\upsilon_{3,k}},~~\forall k\in \mathcal{K}. \tag{\ref{eq2}b}
 \end{align}
 
 We note that only constraint (\ref{eq2}b) is non-convex. Next, we linearize it using the first-order Taylor expansion. Given points $\bar{\upsilon}_{2,k}$ and $\bar{\upsilon}_{3,k}$, the non-convex constraint is convexified as
\begin{equation}
\begin{aligned} \label{eq6}
c_{2,k} & \leq e^{\bar{\upsilon}_{2,k}}\left(\upsilon_{2,k}-\bar{\upsilon}_{2,k}+1\right), \\
2^{r_{k}}-1 & \leq e^{\bar{\upsilon}_{3,k}}\left(\upsilon_{3,k}-\bar{\upsilon}_{3,k}+1\right).
\end{aligned}
\end{equation}

Consequently, problem (\ref{prob33}) is recast as
\begin{align} 
\label{prob34}
\min\limits _{\mathbf{p}, \mathbf{f}_L} & ~~~~\mu_1+\mu_2\\
s.t.
&~~(\ref{prob29}\text{e}),(\ref{prob29}\text{f}),(\ref{prob33}\text{a}),(\ref{eq1}),(\ref{eq2}\text{a}),(\ref{eq3}),(\ref{eq4}),(\ref{eq6}), \notag
\end{align}
which is a convex problem that  can be solved using the standard convex optimization tools such as CVX. By iteratively solving problem (\ref{prob34}) and updating $\bar{\upsilon}_{2,k}$ and $\bar{\upsilon}_{3,k}$, the original problem (\ref{prob32}) can be solved. We outline this approach in Algorithm 3.
\begin{algorithm}[t]
	\setstretch{1}
	\caption{Iterative Optimization based on SCA technique for Solving (\ref{prob32})}
	\begin{algorithmic}[1] 
		\Require  Power budget $P_{max}$, the sub-band allocation matrix $\mathbf{A}$, MEC server CPU frequency $\mathbf{f}_M$, BSs' receive beamforming $\mathbf{u}$, the maximum CPU frequency $F_l$ for vehicles.
		
		\State {Initialize $\bar{\upsilon}_{2,k}$ and $\bar{\upsilon}_{3,k}$.}
		\Repeat
		\State {For given $\bar{\upsilon}_{2,k}$ and $\bar{\upsilon}_{3,k}$, optimize the power control 
			vector $\mathbf{p}$ and  vehicles' local CPU frequency $\mathbf{f}_L$ by solving problem (\ref{prob34}) via SCA technique.}
		\State {Update $\bar{\upsilon}_{2,k}=\upsilon_{2,k}$ and $\bar{\upsilon}_{3,k}=\upsilon_{3,k}$.}
		\Until {The objective value in (\ref{prob34}) convergences.}
		
		\Ensure The transmit power $\mathbf{p}$ and  vehicles' local CPU frequency $\mathbf{f}_L$.
	\end{algorithmic}
\end{algorithm}
\newtheorem{proposition}{Proposition}

\subsection{MEC Server Computation Resource Allocation}
This subsection optimizes the MEC servers' computation resource allocation $ \mathbf{f}_M$ with given sub-band allocation matrix $\mathbf{A}$, the transmit power $\mathbf{p}$, vehicles' CPU frequency  $\mathbf{f}_L$, and BSs' receive beamforming $\mathbf{u}$. The subproblem is written as
\begin{align} 
\label{prob39}
\min\limits _{ \mathbf{f}_M} & ~\max \limits_k ~ \frac{\eta{b}_ke^M_k}{f_{m,k}}\\
s.t.
&~~~(\ref{prob29}\text{d}), \notag
\end{align}
which is a standard convex optimization problem and can be solved via the standard convex optimization tools. However, to reduce computational complexity, we provide the following proposition to obtain  the optimal ${\mathbf{f}_M}$ in closed form.
\begin{proposition}
	The optimal solution of $f_{m,k}^*$ to problem  (\ref{prob39}) is 
	\begin{equation} \label{close2}
	f_{m,k}^*=\frac{\eta{b}_ke^M_k}{\sum_{i=1}^{K_m} \eta{b}_ie^M_i}\hat{F}_m.
	\end{equation}
\end{proposition}
\begin{IEEEproof}
	When other optimization variables are fixed, the sub-problem (\ref{prob39}) can be decoupled for different MEC servers. The computation resource allocation subproblem for MEC server $m$ is given by
	\begin{align} 
	\label{prob40}
	\max\limits _{ \mathbf{f}_M} & ~\min \limits_{k\in \mathcal{K}_m} ~ \frac{\eta{b}_ke^M_k}{f_{m,k}}\\
	s.t.
	&~~~\sum_{k=1}^{K_m}{f}_{m,k} \leq \hat{F}_m. \tag{\ref{prob40}a}
	\end{align}
	
	According to the fairness criterion, when problem (\ref{prob40}) reaches its optimal solution, it follows that $\frac{\eta{b}1e^M_1}{f_{1,k}}=\frac{\eta{b}2e^M_2}{f_{m,2}}=\cdots=\frac{\eta{b}{K_m}e^M{K_m}}{f_{m,{K_m}}}$. Moreover, given $\sum_{k=1}^{K_m}{f}_{m,k} \leq \hat{F}_m$, we have  $f_{m,k}^*=\frac{\eta{b}_ke^M_k}{\sum_{i=1}^{K_m} \eta{b}_ie^M_i}\hat{F}_m$.
\end{IEEEproof}

\subsection{BSs' Receive Beamforming Optimization}
This subsection optimizes BSs' receive beamforming $\mathbf{u}$  with given sub-band allocation matrix $\mathbf{A}$, transmit power $\mathbf{p}$, the CPU frequency of vehicles and MEC servers  $\mathbf{f}_L,\mathbf{f}_M$. The subproblem is given by
\begin{align}
\label{prob35}
\max\limits _{\mathbf{u}} & ~\min \limits_k ~ \frac{R_{k,m}}{\eta{b}_k} \\~~~~~s.t.&~(\ref{prob29}\text{g}).  \notag
\end{align}

It is worth noting that the sub-problem for BSs' receive beamforming can be decoupled for different vehicles. Consequently, the minimum uplink rate maximization problem can be solved individually for each vehicle. Therefore, problem (\ref{prob35}) can be rewritten as follows:
\begin{equation} 
\label{prob36}
{\mathcal{P}}_k^{\mathrm{sub}}({\mathbf{u}}_{k}):\max\limits _{||\mathbf{u}_k||=1}  ~~~ \frac{R_{k,m}}{\eta{b}_k}, ~~\forall k\in \mathcal{K}.
\end{equation}

Since $\log(\cdot)$ is a monotonically increasing function, problem (\ref{prob36}) is equivalent to
\begin{equation} 
\label{prob37}
{\mathcal{P}}_k^{\mathrm{sub}}({\mathbf{u}}_{k}):\max\limits _{||\mathbf{u}_k||=1}  ~~~ \frac{p_k\mathbf{u}^{H}_k\mathbf{h}^{\tilde{l}}_{k,m}(\mathbf{h}^{\tilde{l}}_{k,m})^H\mathbf{u}_k}{\mathbf{u}^H_k\mathbf{D}^{\tilde{l}}_k\mathbf{u}_k}, \forall k\in \mathcal{K}, 
\end{equation}
which is a Rayleigh entropy problem and ${\tilde{l}}$ satisfies $a_{\tilde{l}k} = 1$. Therefore, we have the following proposition

\begin{proposition}
	The optimal solution of $\mathbf{u}_k^*$ to problem (\ref{prob37}) is 
	\begin{equation}\small
	\label{close1}
	\mathbf{u}_k^{*}=\frac{\operatorname{eigvec}\left\{\max \left\{\operatorname{eig}\left\{({\mathbf{D}^{\tilde{l}}_k})^{-1} \mathbf{h}^{\tilde{l}}_{k,m}(\mathbf{h}^{{\tilde{l}}}_{k,m})^H\right\}\right\}\right\}}{\left|\left|\operatorname{eigvec}\left\{\max \left\{\operatorname{eig}\left\{({\mathbf{D}^{\tilde{l}}_k})^{-1} \mathbf{h}^{\tilde{l}}_{k,m}(\mathbf{h}^{{\tilde{l}}}_{k,m})^H\right\}\right\}\right\}\right|\right|}.
	\end{equation}
	where $\operatorname{eig}\{\cdot\}$ denote the eigenvalues and $\operatorname{eigvec}\{\lambda_e\}$ denotes the eigenvector corresponding to the eigenvalue $\lambda_e$.
\end{proposition}
\begin{IEEEproof}
	By defining $\mathbf{w}_k={(\mathbf{D}^{\tilde{l}}_k)}^{1/2}\mathbf{u}_k$, (\ref{prob37}) can be rewriten as
	\begin{equation} \small
	\label{prob38}
	\max\limits _{\mathbf{w}_k}  ~~~ \frac{p_k\mathbf{w}^H_k({\mathbf{D}^{\tilde{l}}_k}^{-1/2}\mathbf{h}_{k,m}^{{\tilde{l}}}(\mathbf{h}^{{{\tilde{l}}}}_{k,m})^H{\mathbf{D}^{\tilde{l}}_k}^{-1/2})\mathbf{w}_k}{\mathbf{w}_k\mathbf{w}^H_k}, \forall k\in \mathcal{K}, 
	\end{equation}
	which is a generalized eigenvector problem \cite{matrix}. Leveraging the properties of Rayleigh entropy, the optimal solution $\mathbf{w}_k^{*}$ can be obtained as the generalized eigenvector corresponding to the largest eigenvalue of the matrix ${\mathbf{D}^{\tilde{l}}_k}^{-1/2}\mathbf{h}_{k,m}^{{\tilde{l}}}(\mathbf{h}^{\tilde{l}}_{k,m})^H{\mathbf{D}^{\tilde{l}}_k}^{-1/2}$. Given that $\mathbf{w}_k={(\mathbf{D}^{\tilde{l}}_k)}^{1/2}\mathbf{u}_k$, the optimal solution $\mathbf{u}_k^{*}$ is the the eigenvector associated with the maximum eigenvalue of the matrix $({\mathbf{D}^{\tilde{l}}_k})^{-1} \mathbf{h}^{\tilde{l}}_{k,m}(\mathbf{h}^{\tilde{l}}_{k,m})^{H}$.
\end{IEEEproof}

\subsection{Overall Algorithm }
\begin{algorithm}[t] 
	\setstretch{1}
	\caption{Overall algorithm for solving problem (\ref{prob29}) }
	\begin{algorithmic}[1] 
		\Require Sensing SINR threshold $\Gamma_{d}$, the computation capacity of MEC server $\hat{F}_m$, the maximum CPU frequency for vehicles $F_l$, maximum iteration number $N^{max}_{it}$.
		
		\State {Initialize  transmit power $\mathbf{p}$, local CPU frequency $\mathbf{f}_L$, MEC server CPU frequency $\mathbf{f}_M$, and BSs' receive beamforming $\mathbf{u}$.}
		\State  {Initialize $\mathbf{A}$ by using  Algorithm 1. }
		\State  {$N_{it} = 1$. }
		\Repeat
		\State {Update $\mathbf{A}$ by using  Algorithm 2. }
		\State {Update  $\mathbf{p}$ and $\mathbf{f}_L$ by using  Algorithm 3.}
		\State {Update $\mathbf{f}_M$ by equation (\ref{close2}).}	
		\State {Update  $\mathbf{u}$ by equation (\ref{close1}).}  	    
		\State {Update $N_{it} = N_{it} + 1$.}
		\Until {$N_{it} \geq N^{max}_{it}$ or the objective value of the problem (\ref{prob29}) converges.}

		\Ensure Sub-band allocation matrix $\mathbf{A}$,  transmit power $\mathbf{p}$,  vehicles' CPU frequency  $\mathbf{f}_L$, BSs' receive beamforming $\mathbf{u}$, and  MEC servers' computation resource allocation  $\mathbf{f}_M$.
	\end{algorithmic}
\end{algorithm}

Building upon the algorithms introduced in the preceding subsections, we now present a unified framework for solving problem (\ref{prob29}) in Algorithm 4. \textcolor{black}{Regarding its convergence, in step 5, note that the update occurs only if the objective value of the newly obtained subchannel allocation solution is smaller than the current objective value.  Besides, $\mathbf{p}$ and $\mathbf{f}_L$ are updated by solving problem (\ref{prob34}) using the SCA technique, and $\mathbf{f}_M$ and $\mathbf{u}$  are updated based on the optimal closed-form solutions. Consequently, the objective value of the problem is non-increasing; hence, the algorithm is guaranteed to converge, as the objective value is lower bounded by 0.}

 The complexity of the overall algorithm primarily arises from Algorithms 1, 2, and 3. In Algorithm 1, the computational complexity mainly comes from  the selection of vehicles and subchannels in lines 6-10, with a complexity of $\mathcal{O}(KL)$. In Algorithm 2 which involves a $K$-level tree, the number of search nodes per level is $LS_{len}$, leading to a complexity of $\mathcal{O}(KLS_{len})$. In Algorithm 3, the complexity is concentrated on solving problem (\ref{prob34}), which can be addressed using the interior-point method with the worst-case computational complexity of $\mathcal{O}(K^{3.5})$. Consequently, omitting lower order terms, the overall complexity of the algorithm 4 is $\mathcal{O}(N_{it}(KLS_{len}+K^{3.5}))$, where $N_{it}$ denotes the number of iterations.

\begin{table}[t]
	\renewcommand{\arraystretch}{1.1}
	\caption{SIMULATION PARAMETERS.}
	\label{table_example2}
	\centering
	\begin{tabular}{l l}
		\hline
		
		\bfseries Parameters &  \multicolumn{1}{c}{\bfseries Value}\\ 
		\hline
		Number of BS antennas  $N$  & 4\\
		Number of BSs $M$  & 4\\
		Number of vehicles $K$  & 48\\
		Number of sub-bands $L$ & 5\\
		Bandwidth of each sub-band $B$  & 10 MHz  \\
		Local computation  intensity  $e_k$ & 50 cycles/bit \\
	    Edge computation  intensity $e_k^M$ & 400 cycles/bit \\
		\tabincell{l}{Maximum CPU frequency \\ for vecicles $F_{l}$}   & 1 Gcycles/s\\
		\tabincell{l}{Computation capacity of\\ MEC server $F_{m}$}   & 30 Gcycles/s\\
		\tabincell{l}{Vehicles' maximum power $P_{max}$} & 30 dBm\\
		The AWGN power at BS $l$ $\sigma_b^u$ & -100 dBm\\
		Power coefficient related to the \\ hardware structure $\kappa$ \cite{UAVMEC}& $10^{-26}$  \\
    	Maximum storage width of each \\ layer for BnB algorithm $S_{len}$ &  $16$ \\	
       Number of symbols accumulated \\for sensing $N_s$ & 500\\
		Sensing SINR threshold of \\ terminals $\Gamma_{d}$ & 20 dB \\
		Minimum safety detection\\ distance $d_{min}$ & 40 m\\
		\hline
		
	\end{tabular}
\end{table}

\section{Numerical Results}
In this section, we demonstrate the effectiveness of the proposed joint optimization design in Algorithm 4. We first outline the simulation setup and comparison schemes, followed by a comprehensive analysis of the simulation results.
\subsection{Simulation Setup and Benchmarks}
\begin{figure}[!t]
	\centering
	\includegraphics[width=3.3in]{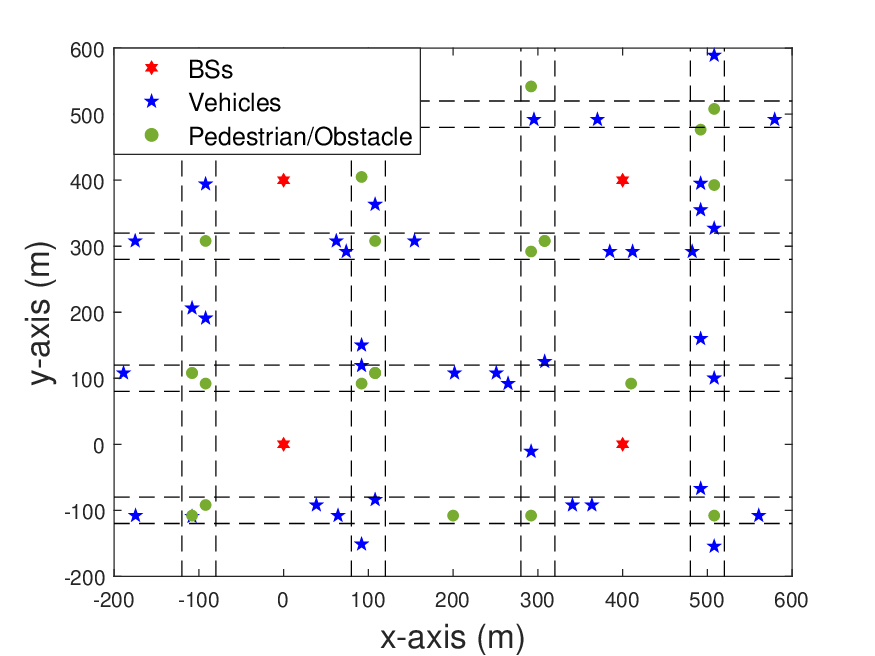}
	\caption{Simulation setup of the  ISCC-V2X network.}
	\label{fig:f3}
\end{figure}

Unless otherwise specified, the following simulation settings are considered. For the ISCC network, as shown in Fig.~\ref{fig:f3}, we consider 4 BSs located within a square area specified by $[-200\text{m}, 600\text{m}]\times[-200\text{m}, 600\text{m}]$, with coordinates at $[0\text{m}, 0\text{m}]$, $[0\text{m}, 400\text{m}]$, $[400\text{m}, 400\text{m}]$, and $[400\text{m}, 400\text{m}]$. Four lanes are arranged equidistantly in both the horizontal and vertical directions, with vehicles randomly distributed across these lanes. We assume that the channels between the BSs and vehicles follow Rayleigh fading, and the RCS of the sensing targets are uniformly distributed between 0.8 and 1. Additionally,  the path loss at reference distance $d_0=1\text{m}$ is set as $\rho_0=-30$ dB. Other key parameters are listed in Table \ref{table_example2}. Furthermore, Monte Carlo simulations are conducted, and the results are averaged over 200 trials to ensure statistical reliability.

To evaluate the performance of the proposed joint optimization design, we consider the following five schemes under the same parameter configurations for comparison.
\begin{itemize}
	\item {\bfseries{Computing-centric resource allocation (CCRA) scheme}}: The CCRA minimizes the completion latency of sensing tasks by optimizing sub-band allocation, power, computation resources, and BSs' receive beamforming, without considering sensing needs of the vehicles. The greedy algorithm-based approach proposed in \cite{kuang2024} is adopted for sub-band allocation, with the optimization of other variables being the same as the joint optimization design. 
	\item {\bfseries{Sensing-centric resource allocation (SCRA) scheme}}: In the SCRA scheme, resource allocation aims to maximize the sensing echoes' SINR rather than task processing latency. BnB algorithm is applied for sensing-centric sub-band allocation. 
	\item {\bfseries{Random sub-band allocation (RSBA) scheme}}: In the RSBA scheme, sub-bands are randomly allocated to different vehicles.
	\item {\bfseries{Fixed power and computation resource (FPCR) scheme}}: In the RPCR scheme, the sub-band allocation and BSs' receive beamforming are jointly optimized.
	\item {\bfseries{Maximum ratio combining (MRC) scheme}}: In this scheme, BSs adopt maximum ratio combining, i.e., $\mathbf{u}_k=\frac{\mathbf{h}_{k,m}}{||\mathbf{h}_{k,m}||}$, to receive signals from different vehicles,   while other resource allocation optimization follows the same steps as the joint optimization design.
\end{itemize}

\begin{figure}[!t]
	\centering
	\includegraphics[width=3.2in]{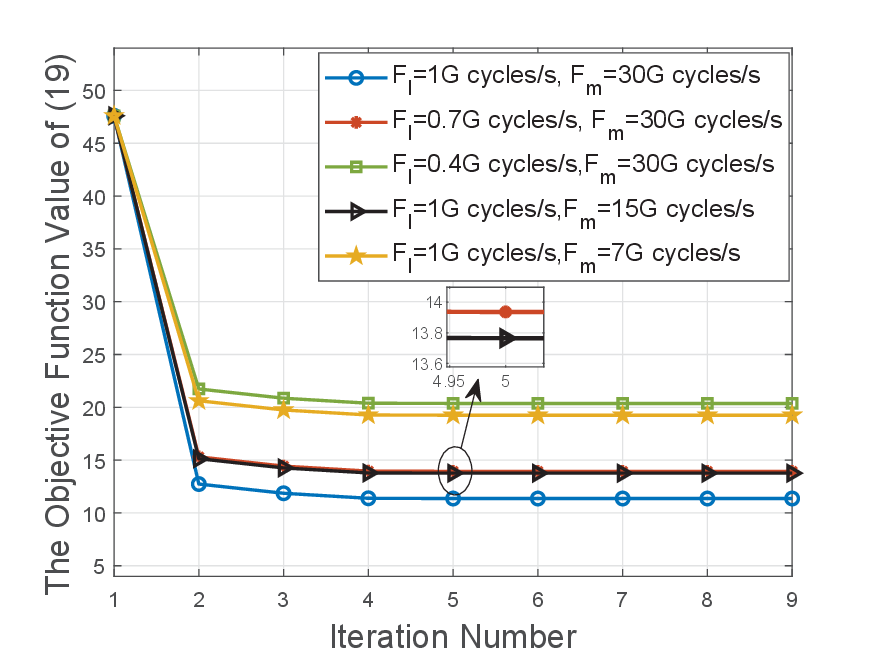}
	\caption{The convergence performance of Algorithm 4.}
	\label{fig:f2}
\end{figure}
\begin{figure*}[t]
	
	\begin{subfigure}
		{	\centering
			\includegraphics[width=2.48in]{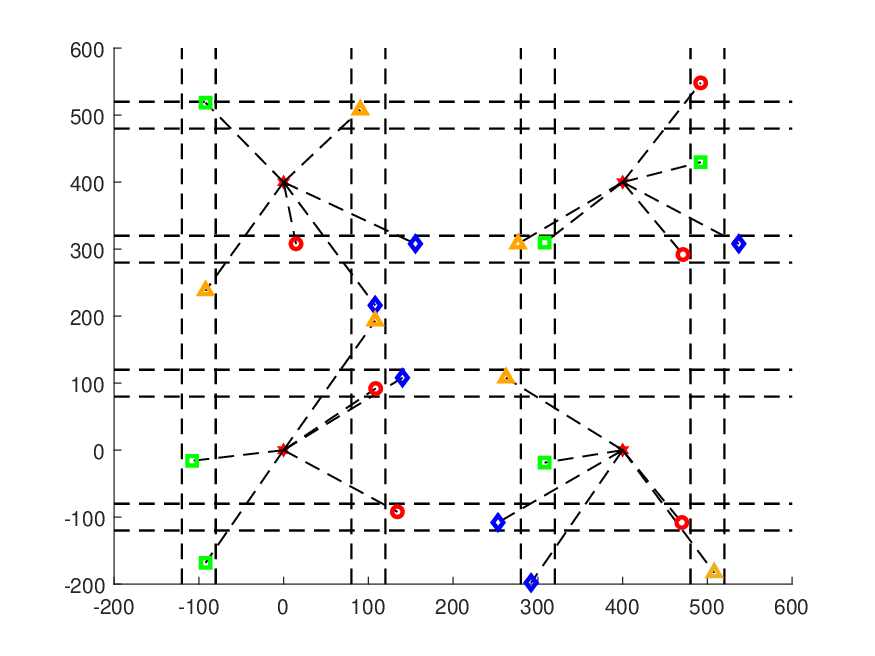}}	
	\end{subfigure}
	\hspace{-7 mm}
	\begin{subfigure}
		{\centering
			\includegraphics[width=2.48in]{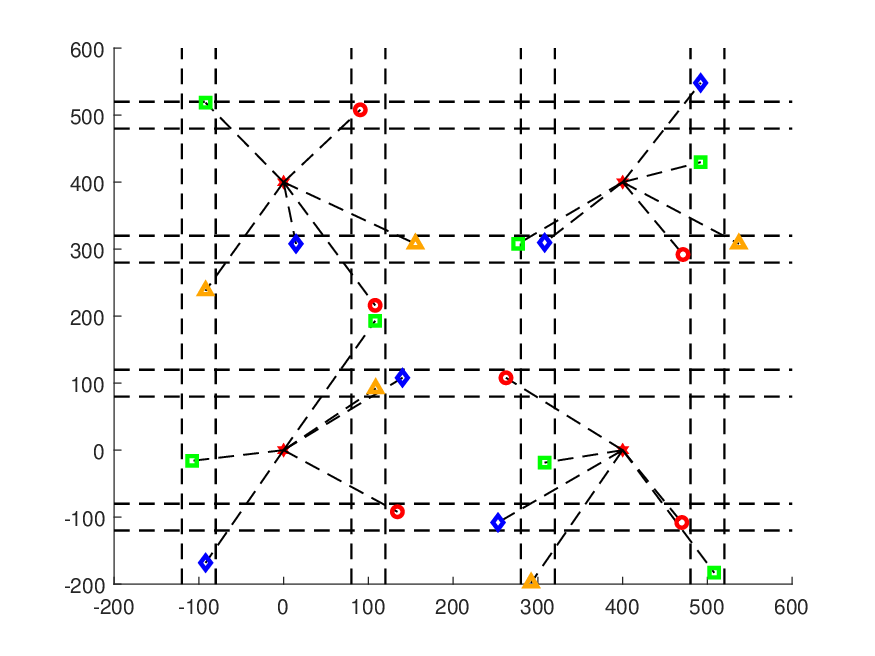}}
	\end{subfigure}
	\hspace{-7 mm}
	\begin{subfigure}
		{\centering
			\includegraphics[width=2.48in]{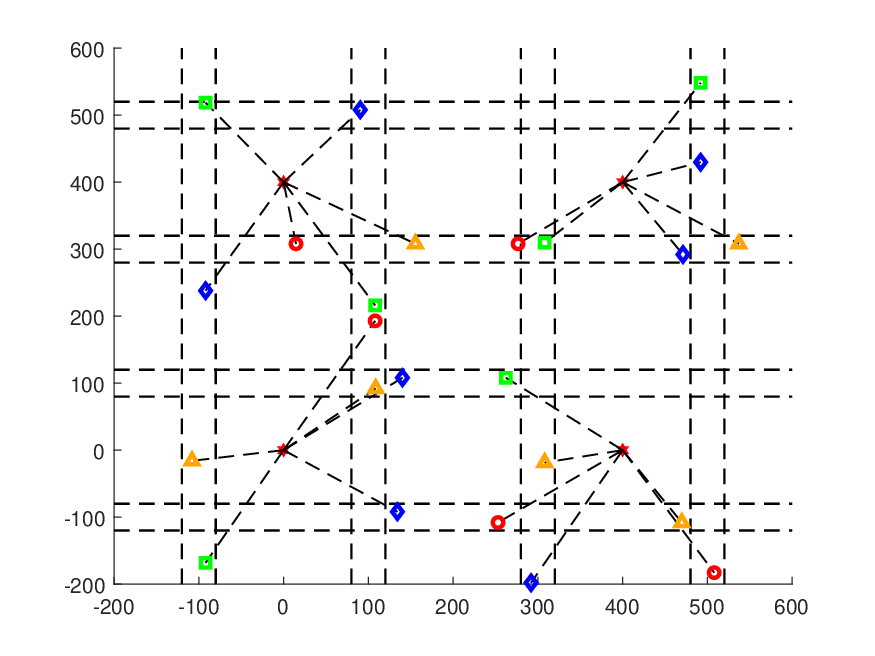}}
	\end{subfigure}
	\caption{(a) Sub-band allocation in the CCRA scheme. (b) Sub-band allocation in the SCRA scheme. (c) Sub-band allocation in the joint optimization design. }
	\label{fig:f1}
\end{figure*}

\subsection{Convergence Performance}
We first  present the convergence performance of our joint optimization design  under different computation resource constraints in Fig.~\ref{fig:f2}. It is seen that the proposed algorithm converges within only a few iterations. As vehicles' available computation resources and MEC severs' computational capacity increase, the maximum processing latency for sensing task decreases accordingly. Moreover, by comparing the red and black lines, we observe that appropriately increasing the computational capacity of MEC servers can achieve similar performance in task completion latency  while requiring fewer local computation resources.
\subsection{Sub-band Allocation Results}
To visually analyze the sub-band allocation characteristics of different schemes, we present the sub-band allocation results with $K=24$ and $L=4$ in Fig. \ref{fig:f1}. Circles of the same color represent vehicles assigned the same sub-band. In the CCRA scheme, we observe that sub-bands are preferentially allocated to vehicles located at similar distances from the BSs to ensure fairness in  task offloading.  This allocation is reasonable because assigning the same sub-band to vehicles with significantly different channel qualities may lead to the strong interference, consequently reducing the data offloading rate for one of the vehicles.  Since the CCRA scheme does not consider sensing, the same sub-band may be allocated to nearby vehicles, resulting in substantial vehicle-to-vehicle interference and negatively impacting sensing performance.  This conclusion may be further illustrated in subsequent analysis.

For the SCRA scheme, which prioritizes sensing performance, sub-bands allocated to neighboring vehicles are always orthogonal to minimize inter-vehicle interference. However, non-orthogonal sub-bands might be assigned to vehicles with significantly different channel qualities, reducing the fairness of task offloading. In contrast, our proposed sub-band allocation scheme not only ensures that sub-bands for neighboring vehicles remain orthogonal but also effectively avoids significant channel quality disparities among vehicles sharing the same sub-band.

\begin{figure}[t]
	\centering
	\includegraphics[width=3in]{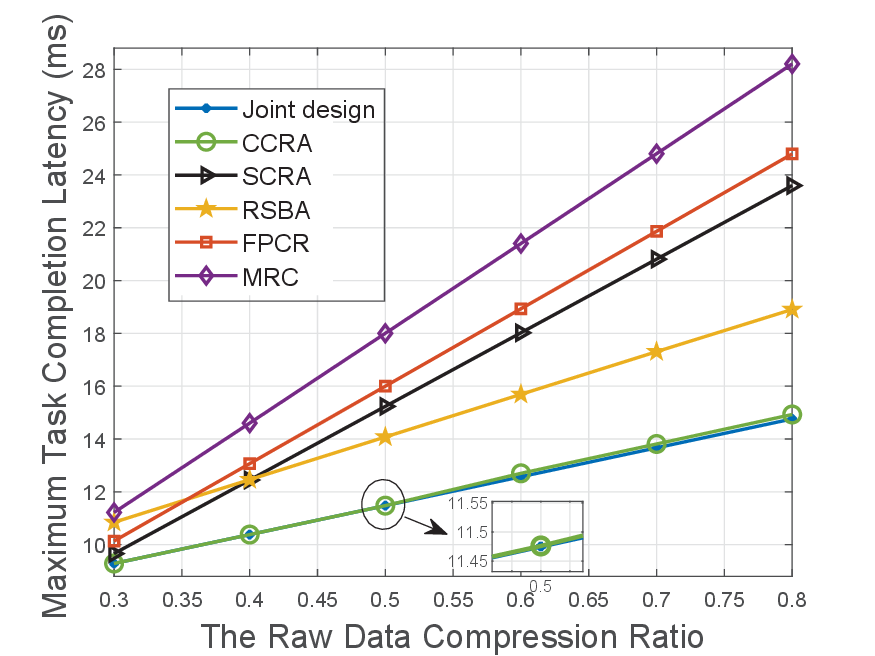}
	\caption{Maximum task completion latency versus different raw data compression ratio $\eta$.}
	\label{fig:f4}
\end{figure}

\subsection{Sensing Task Completion Latency and Target Detection Performance}

Next, we further examine the maximum task completion latency among vehicles and target detection performance of the proposed joint optimization design. Firstly, we show the maximum task completion latency under different schemes in Fig.~\ref{fig:f4}. As can be seen, as the raw data compression ratio $\eta$ increases, the amount of data that needs to be offloaded for processing also increases, leading to a linear increase in both the task offloading latency and the processing latency at the MEC server. Consequently, the maximum task completion latency increases linearly with $\eta$. Compared to RSBA, FPCR, and MRC schemes, the proposed joint optimization design  achieves a lower maximum completion latency, demonstrating the advantages of jointly optimizing sub-bands, computation resources, power, and BSs' receive beamforming. Since SCRA primarily focuses on sensing performance while overlooking the computational process, it exhibits a higher maximum task completion latency compared to the CCRA and joint optimization design. In comparison, although the joint optimization design also considers the sensing performance constraints, its maximum task completion latency is close to that of the CCRA scheme. This is because CCRA employs a greedy-based algorithm \cite{kuang2024} for sub-band allocation, which selects the optimal choice at each step, potentially resulting in a locally optimal solution for the final sub-band allocation. In contrast, the proposed joint optimization design, based on the low-complexity BnB method described in Algorithm 2, explores more potential solutions by continuously branching and pruning,  thereby achieving a sub-band allocation scheme that approximates a globally optimal solution. 

Additionally, in Fig.~\ref{fig:f4}, we also observe that compared to RSBA, CCRA, and the joint optimization design, the latency in SCRA, FPCR, and MRC increase more rapidly with the raw data compression ratio $\eta$. This is because SCRA and FPCR do not optimize local computation resource to mitigate the impact of increased offloading data volumes. Meanwhile, the MRC scheme, due to its ineffective elimination of co-channel interference between vehicles and BSs, results in a much lower offloading rate and is thus more affected by the increase in $\eta$.

\begin{figure}[t]
	\centering
	\includegraphics[width=3in]{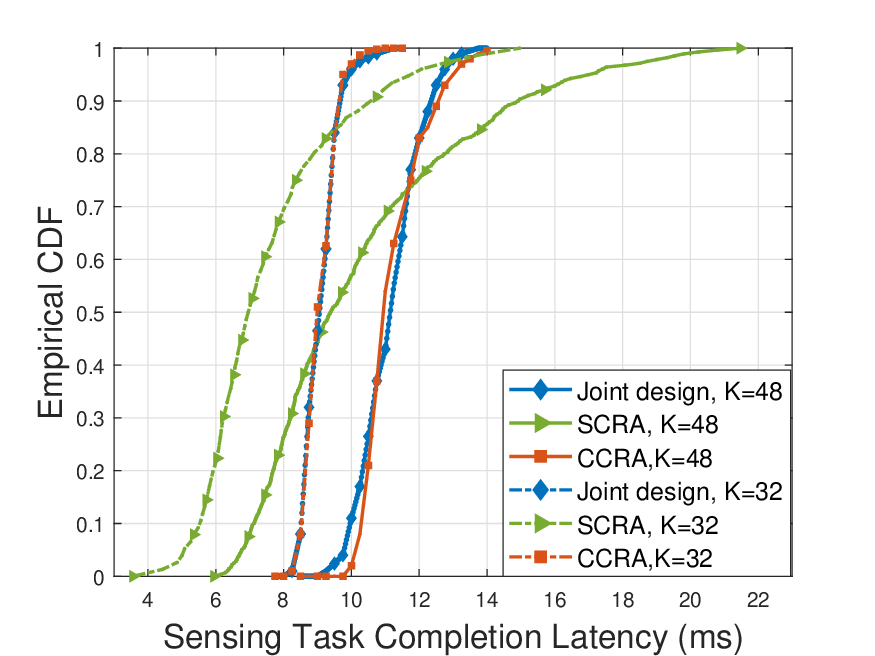}
	\caption{The empirical CDF of vehicles' sensing task completion latency $T_k$.}
	\label{fig:f5}
\end{figure}

In Fig.~\ref{fig:f5}, we present the CDF of vehicles' task completion latency. It is observed that the joint optimization design and CCRA scheme both achieve better fairness in sensing task completion latency compared to the SCRA scheme. This is because they adopt the minimization of the maximum completion latency among vehicles as the optimization objective, while the SCRA scheme focuses on maximizing the sensing echoes' SINR. Additionally, an increase in the number of vehicles $K$ leads to higher maximum task completion latency due to increased vehicle-to-BS interference, which subsequently reduces the offloading rate. Notably, compared to SCRA, the maximum task completion latency reductions of CCRA and joint optimization design grow with the increase of $K$. This is because SCRA prioritizes mitigating vehicle-to-vehicle interference rather than vehicle-to-BS interference; as a result, the increase in vehicle-to-BS interference significantly diminishes the offloading rate and increases the task completion latency.

\begin{figure}[t]
	\centering
	\includegraphics[width=3in]{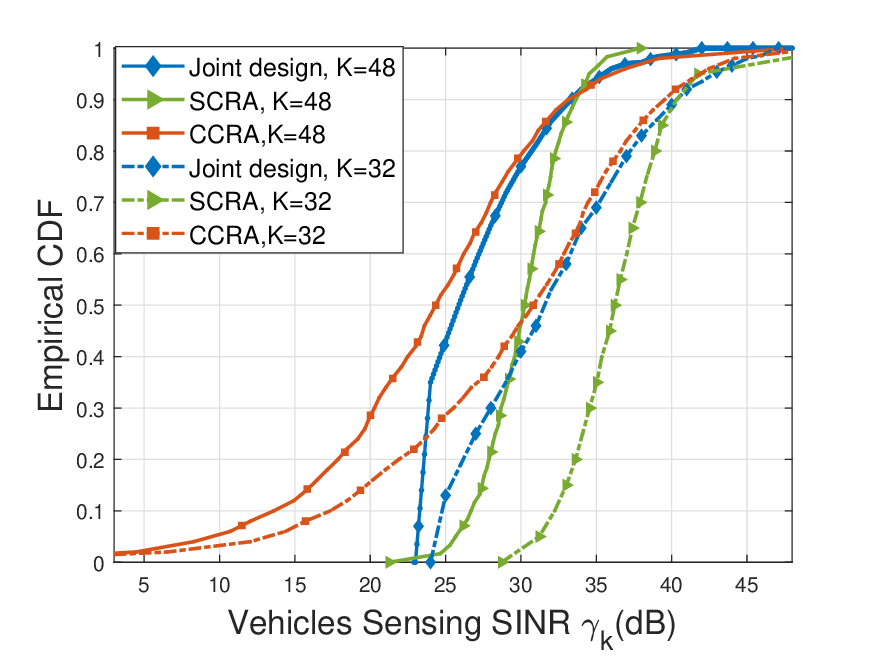}
	\caption{The empirical CDF of vehicles' sensing SINR $\gamma_k$.}
	\label{fig:f6}
\end{figure}

After verifying the effectiveness of the proposed algorithm in reducing  maximum task completion latency, we  evaluate the sensing performance across different schemes. In Fig.~\ref{fig:f6}, we present the CDF of vehicles' sensing SINR for CCRA, SCRA, and joint optimization design schemes. We can see that SCRA exhibits the best overall sensing performance, with over 99\% of vehicles achieving a sensing SINR above 25dB when $K=48$. Moreover, compared to CCRA, the proposed joint optimization design demonstrates a higher sensing SINR, attributed to its dual consideration of sensing task completion latency and vehicle sensing performance in resource allocation design.  Under $\Gamma_{d} = 23$ dB, all vehicles in the joint optimization design achieve a sensing SINR above 23 dB, validating the effectiveness of the sensing constraints. Additionally, we observed that as the number of vehicles $K$ increases, the sensing SINR of all three schemes declines due to increased vehicle-to-vehicle interference.

\begin{figure}[t]
	\centering
	\includegraphics[width=3in]{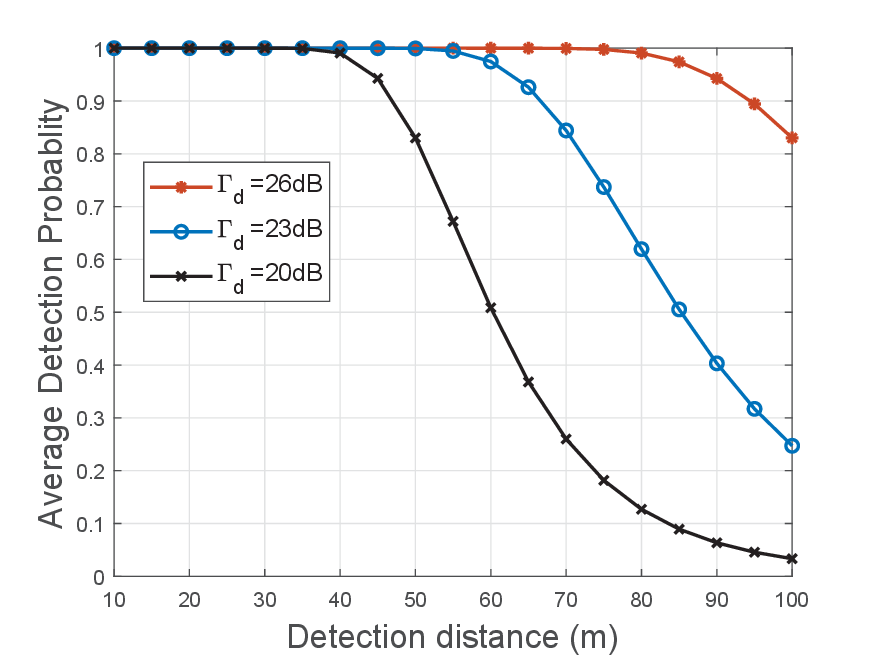}
	\caption{The average detection probability  versus different target distance.}
	\label{fig:f7}
\end{figure}

To more intuitively demonstrate the sensing performance of the joint optimization design, Fig.~\ref{fig:f7} presents the average detection probability under different sensing SINR thresholds with a minimum safety detection distance of $d_{min} = 40$m. It is observed that all three curves achieve a 100\% detection probability at 40m and higher sensing SINR thresholds allow for detection at greater distances, demonstrating the joint optimization design's effectiveness in target detection.

\subsection{Tradeoff between System Performance and Different  Resource Consumption}

\begin{figure}[t]
	\centering
	\includegraphics[width=3in]{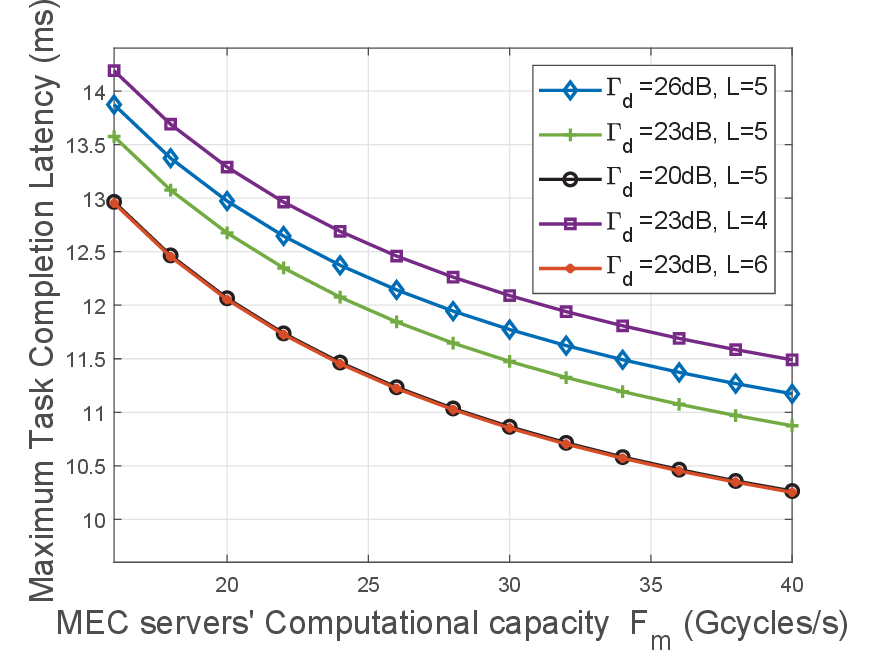}
	\caption{Maximum  task completion latency versus different MEC server's computation capacity $\hat{F}_m$}
	\label{fig:f8}
\end{figure}

\begin{figure}[t]
	\centering
	\includegraphics[width=3in]{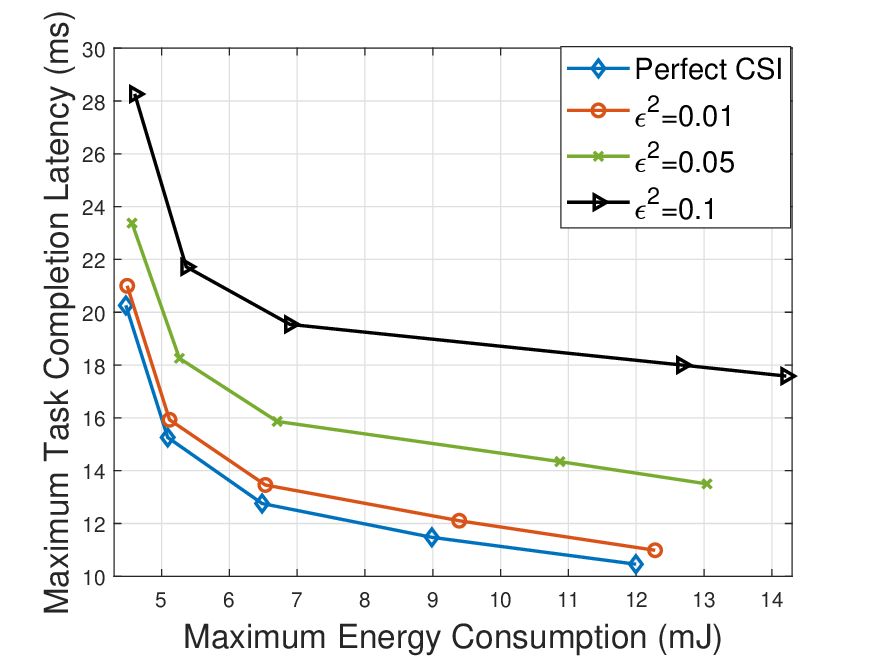}
	\caption{Maximum  task completion latency versus maximum energy consumption.}
	\label{fig:f9}
\end{figure}
		
Finally, we reveal the trade-off between system performance and resource utilization in the ISCC system. Fig.~\ref{fig:f8} shows the maximum task completion latency under various sensing threshold $\Gamma_{d}$ and MEC server computation capacities $\hat{F}_m$. As MEC servers' computational capacity increases, more computation resources are allocated to each vehicle's sensing task, effectively reducing the maximum completion latency. Additionally, we observe that the reduction rate of latency  decreases with increases of $\hat{F}_m$. This is because the maximum task completion latency simultaneously depends on local computing capabilities, offloading rates, and MEC computational capacity. When sufficient MEC servers' computation resources are available, the maximum task completion latency is primarily influenced by local computing capabilities and offloading rates.  

Although we demonstrated in Fig.~\ref{fig:f7} that increasing the sensing threshold $\Gamma_{d}$ effectively  improves the vehicles' target detection performance, Fig.~\ref{fig:f8} shows that as $\Gamma_{d}$ increases, the maximum task completion latency also increases. This is because improved sensing performance necessitates an increase in transmit power to enhance the SINR of target echoes, which leads to increased interference between vehicles and BS, as well as a reduction in local computing \textcolor{black}{capacity}, thereby increasing the  maximum completion latency. Nevertheless, it is observed that increasing the number of sub-bands $L$, effectively reduces the maximum completion latency. Comparing the red and black curves, we find that when the sensing threshold $\Gamma_{d}$ increases, it is possible to maintain the maximum task completion latency unchanged by increasing the number of orthogonal sub-bands, demonstrating the trade-off between sensing performance and sub-band resource utilization. Moreover, it is also observed that by increasing the computation resources of MEC servers, we achieve similar sensing and computing performance with fewer sub-bands, which demonstrates the balance between different resource utilizations.




\textcolor{black}{Fig.~\ref{fig:f9} shows the maximum task completion latency versus the maximum energy consumption.  As the energy consumption increases, the maximum task completion latency decreases. This is because higher energy enables greater transmission power, which improves the offloading rate; at the same time, the available computing frequency at the vehicle also increases, thereby reducing local processing latency. To evaluate the sensitivity of the proposed algorithm to CSI errors, we also present the minimum task completion latency under different CSI error levels in Fig.~\ref{fig:f9}, where $\epsilon^2$ denotes the normalized CSI error. It is observed that when $\epsilon^2=0.01$, the algorithm’s performance degradation is negligible, since small CSI errors do not significantly change the resource allocation strategy. Nevertheless, when the CSI error increases ($\epsilon^2=0.05, 0.1$), the resource allocation becomes less efficient due to inaccurate interference estimation, power control, and beamforming design.}

\begin{figure}
	\setlength{\abovecaptionskip}{-0.1 cm}
	\setlength{\belowcaptionskip}{-1cm}
	\centering
	\begin{subfigure}[]
		{\centering
			\includegraphics[width=1.8in]{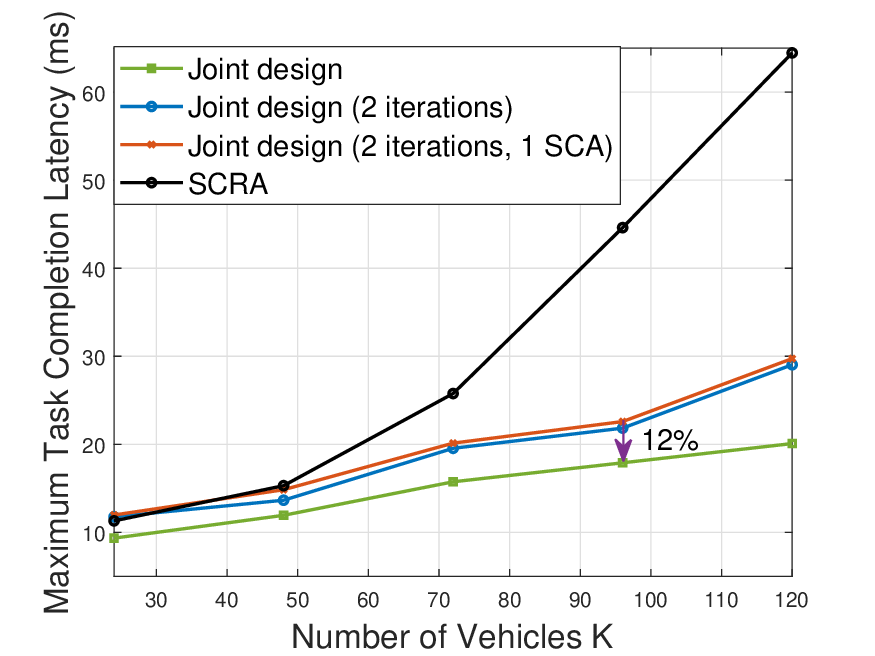}}
	\end{subfigure}
	\hspace{-7 mm}
	\begin{subfigure}[]
		{\centering
			\includegraphics[width=1.8in]{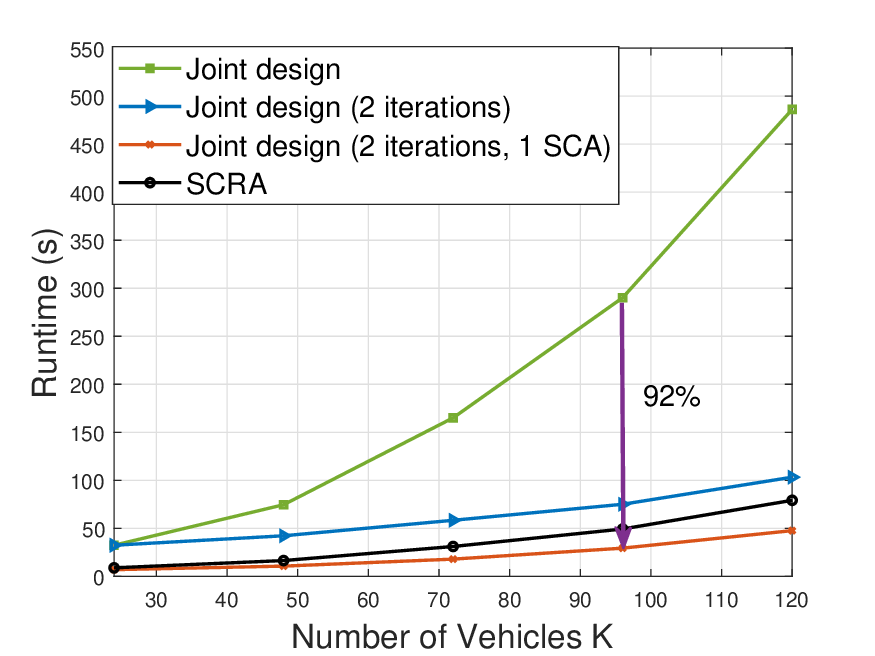}}
	\end{subfigure}
	\caption{(a) Maximum  task completion latency versus the number of vehicles $K$. (b) Algorithm runtime versus the number of vehicles $K$. }
	\label{fig:f10}
\end{figure}

\textcolor{black}{To illustrate the complexity-performance trade-off, Fig. \ref{fig:f10}(a) and Fig. \ref{fig:f10}(b) show  the variations in the maximum task completion latency and algorithm runtime, respectively,  with respect to the number of vehicles $K$. Simulations have been conducted  on an Intel Core i9-13900K processor, with the SeDuMi solver  used to solve the convex problem. We observe that although the proposed joint design scheme achieves lower latency, it also incurs higher computational complexity, which increases with $K$. However, reducing the number of iterations and convex relaxations leads to a 92\% reduction in runtime with only a 12\% performance loss, thereby improving the complexity–performance trade-off and delivering better latency and runtime performance than the SCRA scheme. Furthermore, in practical deployments, optimization can be accelerated on field programmable gate arrays (FPGA) platforms, achieving microsecond-level execution latency \cite{FPGA} and ensuring the algorithm's scalability and feasibility in large systems.}

\section{Conclusion}
In this paper, we investigated the joint optimization of radio and computation resource allocation in an ISCC-aided V2X network, where vehicles offload sensing echoes to MEC servers while detecting targets via unified  ISAC signals. To ensure the fairness of network performance, \textcolor{black}{we aimed to minimize the maximum sensing tasks completion latency among vehicles  by optimizing the sub-bands, power, and computation resource allocation as well as BSs' receive beamforming. An alternating optimization algorithm based on the low-complexity BnB and SCA techniques was proposed.} \textcolor{black}{ Simulation results offer practical insights for the design of ISCC systems. Compared to purely optimizing radio or computation resource allocation, the joint resource allocation scheme significantly reduces sensing task completion latency while ensuring network fairness. Moreover, unlike conventional sub-band allocation schemes that focused solely on either sensing or computation functionality, the proposed algorithm considers both processes, resulting in a better trade-off between sensing and computing performance.}

Although joint radio and computation resource allocation has proven effective for ISCC-enabled V2X systems,  several challenges remain in practical systems. Specifically, in high-mobility scenarios, dynamic propagation environment makes it difficult to obtain perfect CSI. Moreover, the Doppler effect, which induces frequency shifts in communication signals, further complicates CSI estimation and resource allocation. \textcolor{black}{Therefore, our future work aims at addressing the imperfect CSI issues  to enhance both  efficiency and adaptability of resource allocation in highly dynamic environments.} \textcolor{black}{In addition, extending the current framework to incorporate the energy consumption of MEC servers will be considered to enable a more comprehensive system-level energy analysis.}



\bibliographystyle{IEEEtran}
\bibliography{biblp/bibfilelp}

\end{document}